\documentclass[journal, onecolumn, draftcls,12pt]{IEEEtran}
\usepackage{amssymb,amsmath,epsfig,graphicx,theorem}
\usepackage{amsfonts,enumitem}
\usepackage{bm}
\usepackage{arydshln}
\usepackage{algorithm}
\usepackage{algorithmic}
\usepackage{multicol}
\usepackage{caption}

\allowdisplaybreaks

\newtheorem{Theo}{Theorem}
\newtheorem{Lem}{Lemma}
\newtheorem{Cor}{Corollary}
\newtheorem{Def}{Definition}
\newtheorem{Remark}{Remark}

\makeatletter

\newcommand{\Rmnum}[1]{\expandafter\@slowromancap\romannumeral #1@}
\makeatother

\usepackage{epsfig,rotating,setspace,latexsym,amsmath,epsf,amssymb,bm,color}
\usepackage{cite}
\usepackage{tikz}
\usepackage{dblfloatfix}
\usepackage{subfigure}
\usepackage[justification=centering]{caption}

\IEEEoverridecommandlockouts

\title{Capacity Results for the Wiretapped Oblivious Transfer}
\author{Tianyou Pei,  Wei~Kang and Nan Liu%
\thanks{T. Pei and W. Kang are with the School of Information Science and Engineering,
Southeast University, Nanjing, China (email: typei@seu.edu.cn, wkang@seu.edu.cn). 
N. Liu is with the National Mobile Communications Research Laboratory,
Southeast University, Nanjing, China (email:  nanliu@seu.edu.cn). This work is partially supported by the National Natural Science Foundation of China
under Grants  $61971135$ and $62071115$, and the Research Fund of National Mobile Communications Research Laboratory, Southeast University (No. 2022A03).
}
}

\begin{document}

\maketitle

\begin{abstract}
In this paper, we study the problem of the 1-of-2 string oblivious transfer (OT) between Alice and Bob in the presence of a passive eavesdropper Eve. The eavesdropper Eve is not allowed to get any information about the private data of Alice or Bob. When Alice and Bob are honest-but-curious users, we propose a protocol that satisfies $1$-private (neither Alice nor Bob colludes with Eve) OT requirements for the binary erasure symmetric broadcast channel, in which the 
channel provides dependent erasure patterns to Bob and Eve. We find that when the erasure probabilities satisfy certain conditions, the derived lower and upper bounds on the wiretapped OT capacity meet. Our results generalize and improve upon the results on wiretapped OT capacity by Mishra et al. Finally, we propose a protocol for a larger class of wiretapped channels and derive a lower bound on the wiretapped OT capacity. 
\end{abstract}

\section{introduction}

Oblivious transfer (OT) is a fundamental research problem in cryptography. The 1-of-2 string OT involves two parties, commonly named Alice and Bob. Alice has two independent binary strings with the same length, denoted as $K_0$ and $K_1$. Bob is interested in obtaining one of the two strings from Alice, say $K_\Theta$, while keeping his preference $\Theta$ secret from Alice. We call the random variable $\Theta$ as Bob's bit, which takes its value in $\{0,1\}$. At the same time, Alice wishes that Bob knows nothing about the other string, denoted as $K_{\bar{\Theta}}$, where $\bar{\Theta}=1-\Theta$. It has been shown that OT can be achieved with additional noisy resources, either a discrete memoryless multiple source (DMMS) or a noisy discrete memoryless channel (DMC), e.g.,\cite{C1988,C1997,C2004,K2000,S2002}. 

 The concept of OT capacity was introduced by Nascimento and Winter \cite{N2006},\cite{N2008}, which represents the minimum amount of noisy resources, e.g.,~channel uses of the DMC, to achieve OT. Ahlswede and Csisz\'ar derived a general upper bound on the 1-of-2 string OT capacity \cite{C2009}. Furthermore, under the setting that the DMC is a binary erasure channel (BEC) and that the users are honest-but-curious, a protocol is proposed in \cite{C2009}, in which  the existence of a noiseless public channel, where Alice and Bob can communicate for free, is assumed. The noisy BEC channel is utilized in the following way \cite[Theorem 2]{C2009}: Alice sends an i.i.d. binary uniform sequence $X^n$ through the BEC $n$ times. Bob receives $Y^n$ and generates a set $\mathcal{G}$ consisting of the non-erased positions and a set $\mathcal{B}$ of the erased positions and sends $(\mathcal{G},\mathcal{B})$ or $(\mathcal{B},\mathcal{G})$ into the public channel depending on Bob's bit.  
The erasure pattern of Bob, i.e., the positions at which Bob received erasures, is unknown to Alice and therefore is used to conceal Bob's bit. On the other hand, $X_{\mathcal{B}}$ is erased by the BEC channel, and hence, unknown to Bob, and therefore, can be used by Alice to encrypt $K_{\bar{\Theta}}$. It has been shown that in the case of the BEC channel, the lower and upper bounds meet, and the OT capacity is found \cite{C2009}. The protocol that works for the BEC channel can be generalized to a class of erasure-like channels, and the corresponding lower bound on the OT capacity is derived \cite{C2009}. The protocol proposed in \cite{C2009} for the generalized erasure-like channels, when specified to the  BEC model, is different from the protocol specifically designed for the BEC in \cite[Theorem 2]{C2009}. It firstly inflates both $\mathcal{G}$ and $\mathcal{B}$ to the length of $\frac{n}{2}$, which makes one of $\mathcal{G}$ and $\mathcal{B}$ include both erased and non-erased positions, depending the parameter of the BEC. By applying the technique of random binning on both $X_\mathcal{G}$ and $X_\mathcal{B}$, the protocol, though different from the one in \cite[Theorem 2]{C2009}, also achieves the OT capacity in the case of the BEC \cite[Remark 7]{C2009}.

The wiretapped OT problem was introduced by Mishra et al. in \cite{M2017}, which includes an eavesdropper Eve in addition to the original Alice and Bob setting. The legitimate user Bob and the eavesdropper Eve connect to Alice through a broadcast channel, characterized by $p(y,z|x)$, where the input to the channel from Alice is denoted as $X$, and the output received by Bob and Eve are denoted as $Y$ and $Z$, respectively. In addition, Eve can passively receive all the information from the public channel. On top of the OT requirements of Alice and Bob, it is required that Eve can not get any information about the two strings of Alice, nor Bob's bit. 
In this three-party setup, two problems have been proposed \cite{M2017}. One is $1$-privacy, which assumes that neither Alice nor Bob colludes with Eve. The other is $2$-privacy, which assumes that Alice or Bob may collude with Eve to obtain more private information. In this paper, we focus on the $1$-privacy problem.

For the $1$-privacy problem, Mishra et al.\cite{M2017} mainly studied two special classes of broadcast channels. The first class is where the broadcast channel consists of two independent BECs, which we call independent erasure broadcast channel (IEBC), i.e.,  $p(y,z|x)=W(y|x)V(z|x)$, where both $W(y|x)$ and $V(z|x)$ are BECs. For this class, Mishra et al. obtained the wiretapped OT capacity. 
The second class is the physically degraded broadcast channel made up of a cascade of two independent BECs, which we call degraded erasure broadcast channel (DEBC), i.e., Alice connects to Bob via a BEC and Bob connects to Eve via another BEC.  In other words, $p(y,z|x)=W(y|x)V(z|y)$, where both $W(y|x)$ and $V(z|y)$ are  BECs. In this case, they proposed both lower and upper bounds for the OT capacity for $1$-privacy, which do not meet.

For the wiretapped OT problem, under the setting of IEBC, Eve naturally has no knowledge about Bob's bit \cite[Theorem 2]{M2017}, because Eve, like Alice, has no knowledge about the erasure pattern at Bob. 
However, for correlated erasures, such as the DEBC considered in \cite[Theorem 5]{M2017}, Eve can know something about the erasure pattern of Bob from her own erasure pattern, which eventually will leak Bob's bit to Eve. To overcome this problem, Mishra et al. \cite[Theorem 5]{M2017} proposed to utilize the noisy channel to establish some common randomness between Alice and Bob, which is secret from Eve, and  then use this common randomness to encrypt the erasure pattern of Bob. By doing so, Eve will be totally ignorant of the erasure pattern of Bob, and therefore Bob's bit is kept private from Eve. However, a rather large amount of channel resources are needed to generate enough common randomness between Alice and Bob to encrypt the erasure pattern at Bob, and as a result, the protocol in \cite[Theorem 5]{M2017} achieves a lower bound on OT capacity that does not meet with the upper bound. 

Compared to the non-wiretapped OT problem, another difficulty that arises in the wiretapped OT is that, $X_\mathcal{G}$ and/or $X_\mathcal{B}$ can be partially seen by Eve through the BEC channel, which compromises the security of $K_\Theta$ and $K_{\bar{\Theta}}$, as they are encrypted by $X_\mathcal{G}$ and $X_\mathcal{B}$, respectively. To overcome this difficulty, it is proposed in \cite{M2017} that Bob adopts the inflation-binning method, which originated from \cite[Remark 7]{C2009}. In applying the inflation-binning method, \cite{M2017} replaced binning by universal$_2$ hash functions. Under the IBEC and DBEC model considered in \cite{M2017}, we note that the length of $\mathcal{G}$ and $\mathcal{B}$ will not exceed $\frac{n}{2}$. 

In this paper, we focus on the $1$-privacy wiretapped OT problem under the binary erasure symmetric broadcast  channel (BESBC). The set of channels under consideration is more general and contains the IEBC and DEBC studied in \cite{M2017} as special cases. Our novelty compared to the results in \cite{M2017} is two-fold. First, we utilize a 1-bit common randomness between Alice and Bob, which is unknown to Eve, to control the order of Alice's transmission of the encrypted versions of the two strings over the public channel. As a result, Bob's bit is concealed from Eve even when the erasure pattern of Bob is leaked to Eve. Since the cost of establishing a 1 bit common randomness is negligible when the length of the strings are sufficiently large, we find a tighter lower bound on the wiretapped OT capacity. We further show that under certain channel conditions, the proposed lower bound meets the upper bound, and the wiretapped OT capacity is established. Second, given that the BESBC is more general than the IEBC and DEBC, we may encounter the problem where we need to inflate $\mathcal{G}$ and $\mathcal{B}$ to a length larger than $\frac{n}{2}$ before feeding them into the  universal$_2$ hash functions. Note that this difficulty does not exist in the IEBC or DEBC studied in \cite{M2017}. We propose a protocol which reuses part of the non-erased positions at Bob and create some overlap between the sets $\mathcal{G}$ and $\mathcal{B}$. We show that this protocol will not violated the security and privacy constraints due to the observations in \cite[Remark 6]{C2009}. Finally, we utilize the technique of the double random binning and generalize the protocol proposed for the BESBC to the erasure-like  broadcast channels, and obtain the corresponding lower bound on the wiretapped OT capacity.

\section{System Model}

 \begin{figure}[t]
  \centering
  \includegraphics[width=.75\textwidth]{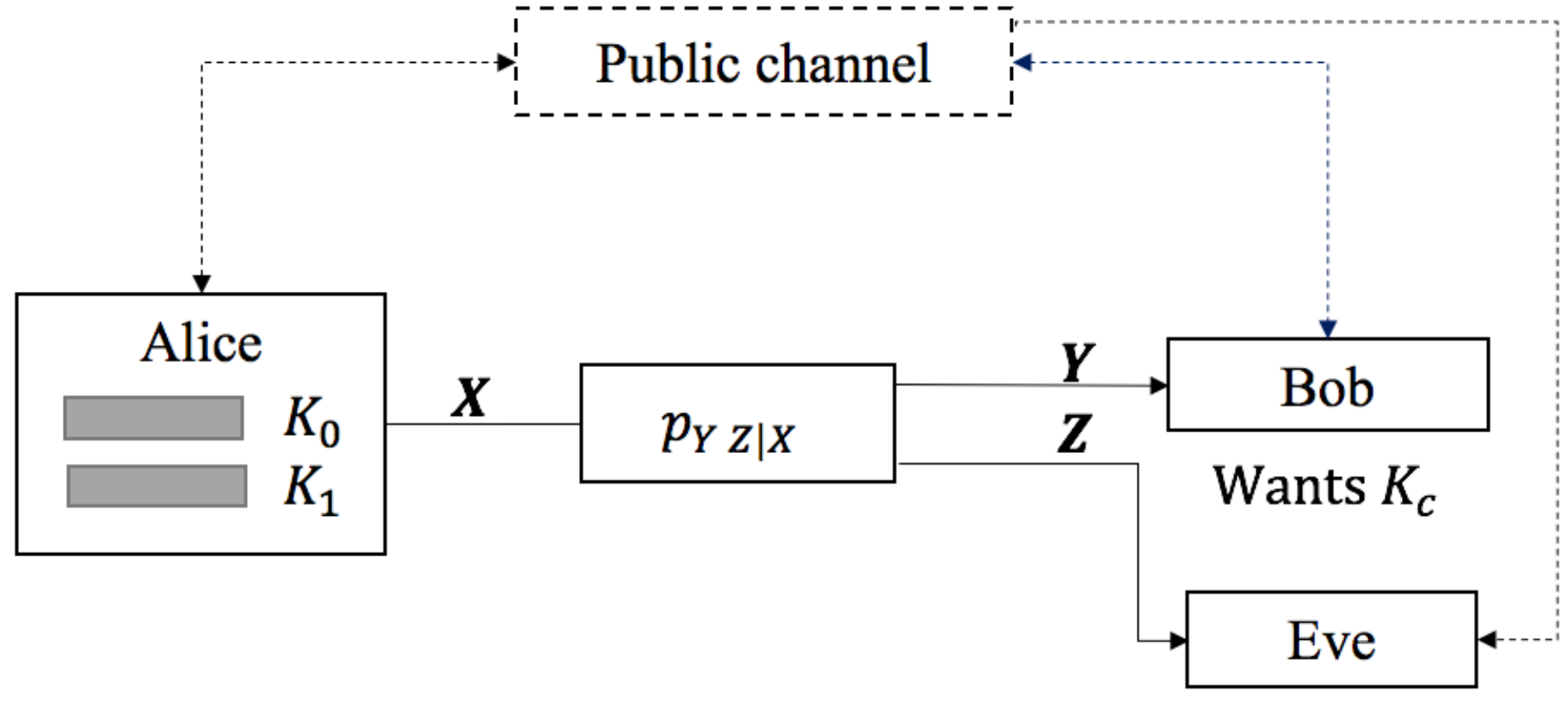}

  \caption{1-of-2 string wiretapped OT} 
  \label{fig:ot_degraded}
\end{figure}

In this paper, we study the $1$-of-$2$ string wiretapped OT problem with $1$-privacy, as shown in Fig~\ref{fig:ot_degraded}.  More specifically, we assume that Alice has two strings with length $k$ bits. The two strings, denoted as $K_0,K_1$, are independent and uniformly distributed. Bob wishes to retrieve $K_\Theta$ from Alice, 
where $\Theta$ is uniformly distributed on $\{0,1\}$, and is denoted as Bob's bit. We also assume that $\Theta$ is statistically independent to $(K_0,K_1)$. 

For the communication between Alice and Bob, the following two constraints for OT need to be satisfied:
\begin{enumerate}
\item Privacy at Alice: Alice learns no information about Bob's bit, i.e., $\Theta$; 
\item Security at Bob: Bob has no information about the unselected string, denoted as $K_{\bar{\Theta}}$, where $\bar{\Theta}=1-\Theta$. 
\end{enumerate}
To achieve OT communication between Alice and Bob, both a public channel, which is noiseless, and a noisy channel $W(y|x)$, where Alice inputs $X$ and Bob receives $Y$, are utilized. We assume that the noiseless public channel is where Alice and Bob can communicate for free. 

In this paper, we further consider the existence of an eavesdropper Eve. Alice connects to Bob and Eve through a wiretap channel, i.e., a noisy channel $p(y,z|x)=W(y|x)V(z|y,x)$, where the input to the channel by Alice is denoted as $X$, and the outputs of the channel for Bob and Eve are denoted as $Y$ and $Z$, respectively. We also assume that Eve can overhear the public channel passively. It is required that Eve can not get any information about the two strings of Alice, i.e., $K_0,K_1$, nor Bob's bit, i.e., $\Theta$.

We assume that Alice, Bob and Eve have unlimited computation power. Furthermore, Alice and Bob can perform random experiments independently of each other to provide the required private randomness. Let random variables $M$ and $N$ represent the private randomness of Alice and Bob respectively. 

\begin{Def}
($k,n$) protocol: 
Alice's private strings $K_0,K_1$ are $k$-bit each.  After initializing the private randomness $M$, Alice transmits a bit $X_i$ over the noisy channel for each channel use, $i=1,...,n$. Before each channel transmission and after the last channel transmission, Alice and Bob send messages alternatively on the noiseless public channel. The messages on the public channel are functions of the user's knowledge, private randomness, the public messages, and channel inputs/outputs the user has seen. The cost of one access to the noisy channel is one unit. Let $F$ denote all messages transmitted on the noiseless public channel at the end of the ($k,n$) protocol. 
\end{Def}

We also assume that Alice and Bob are honest-but-curious users, which means that they comply with the protocol but may infer forbidden information by using available information.
The final knowledge of the two legitimate users and the eavesdropper are
\begin{align}
     V_A & = \{K_0, K_1, X^n, M, F\} \\
V_B & = \{\Theta, Y^n, N, F\} \\
V_E & = \{Z^n, F\}, 
\end{align}
respectively, where  $M$, $N$ are the private randomness generated by Alice and Bob, respectively. Let $\hat{K}_\Theta$, which is a function of $V_B$, denote Bob's estimate for the string of interest $K_\Theta$.  

\begin{Def}
A positive rate $R$ is an achievable wiretapped OT rate with $1$-privacy for honest-but-curious users if for every sufficiently large $n$, every $\tau,\delta>0$, there exists a $(k,n)$ protocol with $\frac{k}{n} \ge R-\delta$, such that
\begin{align}
    \mathsf{Pr}[\hat{K}_\Theta \neq K_\Theta] & \le\tau \label{eqn:ach_1p_wtap_0} \\
I(K_{\bar{\Theta}} ; V_B) & \le \tau \label{eqn:ach_1p_wtap_1} \\
I(\Theta ; V_A) & \le \tau \label{eqn:ach_1p_wtap_2} \\
I(K_0,K_1,\Theta ; V_E) & \le \tau \label{eqn:ach_1p_wtap_3}
\end{align}
\end{Def}
We explain the above conditions as follows: (\ref{eqn:ach_1p_wtap_0}) means that the probability of error at Bob about the string of interest $K_\Theta$ is negligible, (\ref{eqn:ach_1p_wtap_1}) means that Bob remains ignorant about the other message $K_{\bar{\Theta}}$, (\ref{eqn:ach_1p_wtap_2}) means that Alice remains ignorant about Bob's bit, i.e., $\Theta$,  and (\ref{eqn:ach_1p_wtap_3}) means that the eavesdropper Eve can learn nothing about  the two strings of Alice, i.e., $(K_{0},K_{1})$, nor Bob's bit $\Theta$.  Note that we are only interested in $1$-privacy, and as a result, the conditions (\ref{eqn:ach_1p_wtap_1})-(\ref{eqn:ach_1p_wtap_3}) imply that neither Alice or Bob colludes with Eve.

The supremum of achievable wiretapped OT rate is defined as the  wiretapped OT capacity with $1$-privacy, denoted as $C$. 

In this paper, we first focus on the case where the noisy channel $p(y,z|x)=W(y|x)V(z|y,x)$ is a BESBC. More specifically,  
we assume that the input and output alphabets are $\mathcal{X}=\{0,1\}, \mathcal{Y}=\mathcal{Z}=\{0,1,E\}$, where the symbol $E$ denotes the erasure. $W(y|x)$ is a BEC with the probability of erasure being $\epsilon_1$, denoted as BEC($\epsilon_1$). $V(z|y,x)$ is a symmetric channel in the sense that no matter $x=0$ or $1$, $V(z=E|y=E,x)=\epsilon_2$ and $V(z=E|y=x, x)=\epsilon_3$. In other words, $\epsilon_2$ represents the probability that $Z$ is erased given that $Y$ is erased, no matter the input $X$, and $\epsilon_3$ represents the probability that $Z$ is erased given $Y$ is not erased, no matter the input $X$. The above implies that $V(z=x|y=E,x)=1-\epsilon_2$, $V(z =1- x|y=E,x)=0$, $V(z=x|y=x, x)=1-\epsilon_3$, and $V(z=1-x|y=x, x)=0$.

Note that the IEBC and DEBC studied in \cite{M2017} can be viewed as special cases of our model where by setting $\epsilon_2=\epsilon_3$, we obtain the IEBC \cite{M2017}, and by setting $\epsilon_2=1$, we obtain the DEBC \cite{M2017}. 

%

In addition to considering BESBC, we also investigate a more general class of erasure-like broadcast channel as follows. In \cite{C2009}, a more general class of erasure-like memoryless channels are studied, where the channel can be represented as a mixture of two channels with identical input alphabet $\mathcal{X}$ and disjoint output alphabets $\mathcal{Y}_{0}$ and   $\mathcal{Y}_{1}$, namely as
\begin{align}
\label{form:channel of Bob}
    W(y \mid x)=\left\{\begin{array}{r}
(1-\epsilon) W_{0}(y \mid x), \quad x \in \mathcal{X}, y \in \mathcal{Y}_{0} \\
\epsilon W_{1}(y \mid x), \quad x \in \mathcal{X}, y \in \mathcal{Y}_{1} 
\end{array}\right..
\end{align}

In the later sections, we will study the wiretapped OT problem, where the channel from Alice to Bob and the channel from Alice to Eve are both of the form (\ref{form:channel of Bob}) and we have \begin{align}
\mathsf{Pr}[y\in A, z\in B|X=x]=\left\{\begin{array}{cl}(1-\epsilon_1)(1-\epsilon_3)&A=\mathcal{Y}_{0}, B=\mathcal{Z}_{0}\\
(1-\epsilon_1)\epsilon_3&A=\mathcal{Y}_{0}, B=\mathcal{Z}_{1}\\
\epsilon_1(1-\epsilon_2)&A=\mathcal{Y}_{1}, B=\mathcal{Z}_{0}\\
\epsilon_1\epsilon_2&A=\mathcal{Y}_{1}, B=\mathcal{Z}_{1}
\end{array}
\right. .\label{Nan04}
\end{align}

\section{Capacity Results for the wiretapped OT for the BESBC}
\subsection{Main Results}
We first propose a converse result in the following theorem, which provides an upper bound on the capacity of the wiretapped OT in the case of a BESBC.

\begin{Theo}
\label{Theo:upper bound}
The  wiretapped OT capacity with $1$-privacy  for honest-but-curious users for the BESBC satisfies
\begin{align}
    C \leq \min\left\{\epsilon_3(1-\epsilon_1), \epsilon_1, \frac{1}{2}(\epsilon_1 \epsilon_2+ \epsilon_3(1-\epsilon_1)) \right\}.
\end{align}
\end{Theo}

\begin{IEEEproof}
Theorem \ref{Theo:upper bound} will be proven in Section \ref{Nan01}. The proof follows similar ideas as those in \cite[Theorem 1]{C2009} and \cite[Lemma 4]{M2017}.
\end{IEEEproof}

The next theorem states an achievability result, which is a lower bound on the wiretapped OT capacity of the BESBC. However it is only valid for $\epsilon_2 \ge \epsilon_3$.

\begin{Theo}
\label{Theo:lower bound}
A lower bound on the wiretapped OT capacity with $1$-privacy for honest-but-curious users for the BESBC is 
\begin{align}
    C \ge \min\left\{\epsilon_3(1-\epsilon_1), \epsilon_1, \frac{1}{2}(\epsilon_1 \epsilon_2+ \epsilon_3(1-\epsilon_1)) \right\}
\end{align}
if $\epsilon_2 \ge \epsilon_3$.
\end{Theo}
\begin{IEEEproof}
Theorem \ref{Theo:lower bound} will be rigorously proven in Section \ref{Nan02}. We note the general idea here. To motivate our scheme, we first briefly recall the existing achievability schemes proposed for the OT under the setting of the BEC by  Ahlswede and Csisz\'ar\cite{C2009} and the wiretapped OT under the setting of the IEBC and DEBC by Mishra et al. \cite[Theorem 5]{M2017}. 

We begin with the scheme by Ahlswede and Csisz\'ar\cite{C2009}. Alice sends an i.i.d. binary uniform sequence $X^n$ through a binary erasure channel. Bob receives the sequence perfectly in some positions and erasures in others. Bob forms a set $\mathcal{G} \subset \{1,\cdots,n\}$ which is a subset of the positions of the non-erased $Y^n$,  and another set $\mathcal{B} \subset \{1,\cdots,n\}$ which is a subset of the positions of the erased $Y^n$. It is required that $|\mathcal{G}|=|\mathcal{B}|$ and they are both equal to the length of the strings at Alice, i.e., $k$. If Bob wants to retrieve $K_0$, then  define $(\mathcal{L}_0,\mathcal{L}_1)=(\mathcal{G}, \mathcal{B})$, otherwise define $(\mathcal{L}_0,\mathcal{L}_1)=(\mathcal{B},\mathcal{G})$.  Bob sends $(\mathcal{L}_0,\mathcal{L}_1)$ to Alice via the noiseless public channel. Alice retrieves the corresponding parts from $X^n$, 
i.e.,~$X_{\mathcal{L}_0},X_{\mathcal{L}_1}$. These two parts are used as secret keys to encrypt the strings $K_0,K_1$, and then Alice sends $K_0\oplus X_{\mathcal{L}_0}$ and $K_1\oplus X_{\mathcal{L}_1}$ to Bob via the public channel. The privacy of Bob's bit at Alice hinges on the fact that Alice has no knowledge about the erasure pattern, i.e., $\mathcal{B}$ or $\mathcal{G}$, at Bob, and therefore could not tell whether $(\mathcal{L}_0,\mathcal{L}_1)=(\mathcal{G},\mathcal{B})$ or $(\mathcal{L}_0,\mathcal{L}_1)=(\mathcal{B},\mathcal{G})$. Security at Bob is guaranteed because Bob only has the knowledge of $X_{\mathcal{G}}$ and no knowledge at all about $X_{\mathcal{B}}$. 

For the wiretapped OT problem, in IEBC, i.e.,~$\epsilon_2=\epsilon_3$, Eve naturally has no knowledge about Bob's bit,  because Eve, like Alice, has no knowledge about the erasure pattern at Bob. However, for correlated erasures, i.e., $\epsilon_2\ne\epsilon_3$, such as the DEBC studied in \cite[Theorem 2]{M2017} with $\epsilon_2=1$, Eve can determine $(\mathcal{L}_0,\mathcal{L}_1)=(\mathcal{G},\mathcal{B})$ or $(\mathcal{L}_0,\mathcal{L}_1)=(\mathcal{B},\mathcal{G})$ with certainty if the the sequence length $n$ is sufficiently large, and this leaks Bob's bit $\Theta$ to Eve. To cope with this problem, Mishra et al. \cite[Theorem 5]{M2017} proposed to utilize the noisy channel to establish some common randomness between Alice and Bob, which is secret from Eve, and  then use this common randomness to encrypt $(\mathcal{L}_0, \mathcal{L}_1)$. By doing so, Eve will be totally ignorant of $(\mathcal{L}_0, \mathcal{L}_1)$, and therefore can not tell if it is $(\mathcal{G},\mathcal{B})$ or $(\mathcal{B},\mathcal{G})$. Thus, Bob's bit is kept private from Eve.

We note the following two points in the protocol designed in \cite[Theorem 5]{M2017}: 
\begin{enumerate}
\item It requires a large amount of channel resources, i.e.,  $2|\mathcal{L}_0|$ bits of common information between Alice and Bob to generate two secret keys to encrypt $\mathcal{L}_0$ and $\mathcal{L}_1$. As a result, the rate of the protocol in \cite[Theorem 5]{M2017} does not meet the upper bound. 
\item Fundamentally, it is only required that Eve be ignorant of Bob's bit $\Theta$, but not the values of $(\mathcal{L}_0,\mathcal{L}_1)$. 
\end{enumerate}
Therefore, in this paper, we design a protocol that allows Eve to gain the knowledge of $(\mathcal{L}_0, \mathcal{L}_1)$ 
but prevents Eve from learning the value of Bob's bit $\Theta$. 

We note that in the schemes by both Ahlswede and Csisz\'ar\cite{C2009} and  Mishra et al. \cite[Theorem 5]{M2017}, Alice is supposed to send the encrypted versions of $K_0,K_1$ into the public channel in the fixed order, i.e, $K_0$ first and then $K_1$. This fixed order directly connects $(\mathcal{G},\mathcal{B})$ and Bob's bit $\Theta$. More specifically, if we know $(\mathcal{L}_0,\mathcal{L}_1)=(\mathcal{G},\mathcal{B})$, then we know $\Theta=0$; otherwise,  $\Theta=1$. With the above understanding, we propose to change the fixed order to a random order. We design a binary random variable $S$ to control the order. If $S=0$, we send the encrypted version of $K_0,K_1$ in this order into the public channel, and if $S=1$, we send encrypted $K_1,K_0$ in this order into the public channel. As long as Bob knows the value of $S$, the scheme still works. At the same time, Eve knows nothing about $\Theta$ even with the knowledge of $(\mathcal{L}_0, \mathcal{L}_1)$ equal to $(\mathcal{G},\mathcal{B})$ or $(\mathcal{B},\mathcal{G})$. It is obvious that, in this setting, we only need to establish $1$ bit of common randomness between Alice and Bob, that is secret to Eve. The burden of doing so is negligible in terms of the rate.

Another problem in the wiretapped OT is that with the presence of eavesdropper Eve, Alice and Bob need to use universal$_2$ hash functions (or random binning from the information theoretic perspective) to establish a common randomness, which is secret to Eve, to encrypt $K_\Theta$. To establish a common randomness with length $nR$, we need to form sets $\mathcal{G}$ and $\mathcal{B}$ from the non-erased and erased coordinates with length $n\beta=\frac{nR}{\epsilon_3}$. When $\beta>\frac{1}{2}$, we face the problem of not having enough separate coordinates to place into the sets $\mathcal{G}$ and $B$. Note that $\beta>\frac{1}{2}$ will not occur in the IEBC or DEBC studied in \cite{M2017}, i.e.,~$\epsilon_2=\epsilon_3$, or $\epsilon_2=1$. However, when we generalize the model to $\epsilon_2\ge\epsilon_3$, we will face this difficulty. As pointed out in \cite[Remark 6]{C2009}, the correlation between $X_\mathcal{G}$ and $X_\mathcal{B}$  does not matter as long as $X_\mathcal{G}$ has the identical distribution as  $X_\mathcal{B}$. Therefore, we propose to reuse part of the non-erased coordinates in the set $\mathcal{B}$, i.e., to have some overlap between the sets $\mathcal{G}$ and $\mathcal{B}$, which will not violated the security and privacy constraints and make the scheme work at the same time. The details of the proof of Theorem \ref{Theo:lower bound} is in Section \ref{Nan02}.
\end{IEEEproof}


It is straightforward to see that the upper bound in Theorem \ref{Theo:upper bound} and the lower bound in Theorem \ref{Theo:lower bound} meet if $\epsilon_2 \ge \epsilon_3$, which means that we have obtained the wiretapped OT capacity of the BESBC under the condition $\epsilon_2 \ge \epsilon_3$.    


\begin{Remark}
\normalfont We can see that the condition $\epsilon_2 \ge \epsilon_3$, under which we have obtained the OT capacity, includes the two classes studied in \cite{M2017}:
\begin{enumerate}    
    \item In the case of IEBC, i.e.,~$\epsilon_2 = \epsilon_3$, our result collapses to \cite[Theorem 2]{M2017}.
    \item In the case of DEBC, i.e.,~$\epsilon_2=1$, our capacity result coincides with the upper bound given in \cite[Theorem 5]{M2017}, and is strictly better than the lower bound in \cite[Theorem 5]{M2017}.
\end{enumerate}
Therefore, the improvement of our capacity results over the results in \cite{M2017} are:
\begin{enumerate}
\item We propose an upper bound on the wiretapped OT capacity for general $\epsilon_1, \epsilon_2$ and $\epsilon_3$, while \cite{M2017} has an upper bound on the wiretapped OT capacity in the two special cases of   $\epsilon_2 = \epsilon_3$ and $\epsilon_2=1$.
\item We propose a lower bound on the wiretapped OT capacity when $\epsilon_2 \geq \epsilon_3$, while \cite{M2017} has lower bounds on the wiretapped OT capacity in the two special cases of  $\epsilon_2=1$ and $\epsilon_2 = \epsilon_3$. 
\item For $\epsilon_2=1$: we propose an achievable scheme that performs strictly better than the achievable rate in \cite{M2017}. Furthermore, this achievable scheme is shown to be optimal, i.e., capacity-achieving. 
\item For $1>\epsilon_2>\epsilon_3$: we found the capacity of this case, which is not studied in \cite{M2017}.
\end{enumerate}
\end{Remark} 



\subsection{Proof of Theorem 1} \label{Nan01}
To prove Theorem~\ref{Theo:upper bound},  we need to prove that the $1$-private OT rate $R$ for the binary broadcast symmetric erasure channel is bounded above by
\begin{align}
\min \left\{ (1-\epsilon_1)\epsilon_3,\epsilon_1,\frac{1}{2}(\epsilon_1 \epsilon_2+ \epsilon_3(1-\epsilon_1))\right\}
\end{align}

We begin with the first term. If we assume the existence of a genie, who informs Alice the value of $\Theta$ and the erasure patterns, i.e., the set of erasure positions, at both Bob and  Eve, then the problem of OT is equivalent to establish common information between Alice and Bob, which is secret to Eve, i.e., $K_\Theta$. The common information can only be delivered to Bob at the coordinates which is unerased at Bob but erased at Eve. Hence, the rate is upper bounded by $(1-\epsilon_1)\epsilon_3$.

For the second term, we follow the same steps as those in proving equation (16) in \cite[Theorem 1]{C2009} and we can show that the rate is upper bounded by 
\begin{align}
R\le\frac{1}{n}\sum_{t=1}^n H(X_t|Y_t)\le\epsilon_1
\end{align}

%

For the third term, we follow the same proof steps as those in \cite[Lemma 4]{M2017} as follows
\begin{align}
    2nR&=H(K_0,K_1)\nonumber\\
    &=I(K_0,K_1;X^n,F)+H(K_0,K_1|X^n,F)\nonumber\\
    &\overset{(a)}{\le}I(K_0,K_1;X^n,F)+\tau\nonumber\\
    &\leq I(K_0,K_1;X^n,Z^n,F)+\tau\nonumber\\
    &=I(K_0,K_1;Z^n,F)+I(K_0,K_1;X^n|Z^n,F)+\tau\nonumber\\
    &\overset{(b)}{=}I(K_0,K_1;X^n|Z^n,F)+2\tau\nonumber\\
    &=H(X^n|Z^n,F)-H(X^n|Z^n,F,K_0,K_1)+2\tau\nonumber\\
    &\leq H(X^n|Z^n,F)+2\tau\nonumber\\
    &\leq H(X^n|Z^n)+2\tau\nonumber\\
    &\leq n(\epsilon_1 \epsilon_2+ \epsilon_3(1-\epsilon_1))+2\tau
\end{align}
where (a) is due to \cite[Lemma 5]{M2017}, and (b) is because of the condition (\ref{eqn:ach_1p_wtap_3}).

\subsection{Proof of Theorem 2} \label{Nan02}

We present our protocol in details as follows. Alice sends an i.i.d binary uniform sequence  $X^n$ into the broadcast channel. Bob observes $Y^n$ and Eve observes $Z^n$. 
Alice and Bob establish a $1$-bit common randomness as follows. Bob randomly picks $n\alpha$ bits from the non-erased coordinates, called $\mathcal{L}_\alpha$. Then, Bob uniformly pick a function $F_{\alpha}$ from a family $\mathcal{F}$ of universal$_2$ hash functions\cite{Carter77universalclasses} (see Appendix A for details):
\begin{align} F_\alpha : \{0,1\}^{\alpha n} \longrightarrow \{0,1\}^{(\alpha (\epsilon_{3}-\delta)-\bar{\delta})n }  
\end{align}
    where $\delta \in (0,1)$, $(\epsilon_3 - \delta) \in \mathbb{Q}$ and $\bar{\delta} \in (0, \alpha \epsilon_3) \cap \mathbb{Q}$. Bob applies the function $F_{\alpha}$ on $X_{\mathcal{L}_\alpha}$ and retrieves the first bit of $F_\alpha(X_{\mathcal{L}_\alpha})$, denoted by $K_\alpha$, and independently generates a uniform bit $S$. Bob sends $\mathcal{L}_\alpha, F_\alpha, K_\alpha\oplus S$ into the public channel. Alice can recover $S$, while $S$ remains nearly independent to Eve.

Next, Bob will forms two sets $\mathcal{G}$ and $\mathcal{B}$ as follows. Define $\beta=\frac{R}{\epsilon_3}$, where
\begin{align}
R< \min\left\{ \epsilon_3(1-\epsilon_1), \epsilon_1,\frac{1}{2}(\epsilon_1 \epsilon_2+ \epsilon_3(1-\epsilon_1)) \right\}
\end{align} 
Bob uniformly selects $n\beta$ coordinates from the non-erased coordinates except $X_{\mathcal{L}_\alpha}$  and names the set of the corresponding locations as $\mathcal{G}$. 
Next Bob will form the set $\mathcal{B}$, which falls into one of the following three cases:
\begin{enumerate}
\item If $\beta\le\epsilon_1$ and $\beta<\frac{1}{2}$, Bob uniformly selects $n\beta$ coordinates from the erased coordinates and name the set of corresponding locations as $\mathcal{B}$. The choice of $\mathcal{B}$ in this case is similar to \cite[Theorem 2]{C2009}. 
\item If $\epsilon_1\le \beta<\frac{1}{2}$, Bob collects all erased coordinates and uniformly selects $n(\beta-\epsilon_1)$ coordinates from the non-erased coordinates, but not in $\mathcal{G}$ or $\mathcal{L}_\alpha$. Bob names the set of corresponding locations as $\mathcal{B}$. The choice of $\mathcal{B}$ in this case is similar to \cite[Remark 7]{C2009} and \cite[Theorem 1]{M2017}.
\item If $\beta\ge\frac{1}{2}$, Bob collects all the coordinates not in $\mathcal{G}$ or $\mathcal{L}_\alpha$  and uniformly selects $n(2\beta-1)$ coordinates from $\mathcal{G}$. Bob names the set of corresponding locations as $\mathcal{B}$. The choice of $\mathcal{B}$ in this case is a novel step that we propose in this paper.
\end{enumerate} 
After forming the sets $\mathcal{G}$ and $\mathcal{B}$, Bob checks the value of $\Theta$ and $S$, or more specifically the value of $\Theta\oplus S$, and define
\begin{align}
(\mathcal{L}_0,\mathcal{L}_1)=\left\{\begin{array}{rl}(\mathcal{G},\mathcal{B})&\text{ if }\Theta=0, S=0 \text{ or }\Theta=1,S=1, \text{ i.e., }\Theta\oplus S=0\\(\mathcal{B},\mathcal{G})&\text{ if }\Theta=0, S=1 \text{ or }\Theta=1,S=0, \text{ i.e., }\Theta\oplus S=1
\end{array}\right.
\end{align}
or more compactly, 
\begin{align}
\mathcal{G}&=\mathcal{L}_{\Theta\oplus S}\\
\mathcal{B}&=\mathcal{L}_{\bar{\Theta}\oplus S}
\end{align}
Bob then sends $(\mathcal{L}_0,\mathcal{L}_1)$ to Alice through the public channel. This step is the same as \cite{C2009} and \cite{M2017}.

Alice randomly and independently chooses functions $F_0,F_1$ from a family $\mathcal{F}$ of universal$_2$ hash functions:
	\begin{align} F_0,F_1 : \{0,1\}^{\beta n} \longrightarrow \{0,1\}^{n(R - \tilde{\delta})}\end{align}
where $\tilde{\delta} \in (0,R)$. Alice then applies $F_0$ on $X_{\mathcal{L}_0}$ and $F_1$ on $X_{\mathcal{L}_1}$. Finally, Alice checks the value of the random variable $S$. 
\begin{enumerate}
\item If $S=0$, then send $F_0,F_1,K_0\oplus F_0(X_{\mathcal{L}_0}),K_1\oplus F_1(X_{\mathcal{L}_1})$ into the public channel in this order. 
\item If $S=1$, then send $F_0,F_1,K_1\oplus F_0(X_{\mathcal{L}_0}),K_0\oplus F_1(X_{\mathcal{L}_1})$ into the public channel in this order. 
\end{enumerate}
This final step is different from \cite{C2009} and \cite{M2017}, where Alice always sends $K_0$ first and then $K_1$. We use the random variable $S$ to control the sending order to conceal the knowledge of $\Theta$ from Eve.

\begin{Lem}
\label{lem:C1p}
Any rate $R < \min \left\{\epsilon_3(1-\epsilon_1), \epsilon_1,\frac{1}{2}(\epsilon_1 \epsilon_2+ \epsilon_3(1-\epsilon_1))\right\}$ is an achievable wiretapped OT rate with 1-privacy for the BESBC where the users are honest-but-curious and $\epsilon_2\ge \epsilon_3$.
\end{Lem}

Lemma \ref{lem:C1p} is proved in Appendix \ref{Nan03}. 

\section{Extensions Beyond BESBC}
In the previous section, we focus on the wiretapped OT problem for BESBC. In this section, we will follow \cite[Theorem 4]{C2009} and generalize our results to a larger class of broadcast channels as defined in (\ref{Nan04}).

We first define the following quantities:
\begin{align}
C_0 = I(X;Y_0) - I(X;Y_1)\\
C_{11} = I(X;Y_0) - I(X;Z_0)\\
C_{12} = I(X;Y_0) - I(X;Z_1)\\
C_{21} = I(X;Y_1) - I(X;Z_0)\\
C_{22} = I(X;Y_1) - I(X;Z_1)\\
C_G=(1-\epsilon_3)C_{11}+\epsilon_3C_{12}\\
C_B=(1-\epsilon_2)C_{21}+\epsilon_2C_{22}\\
C_N=(1-\epsilon_2)C_{11}+\epsilon_2C_{12}=C_B+C_0
\end{align}
Then, we have the following theorem. 

\begin{Theo}
\label{Theo:general lower bound}
A lower bound on the wiretapped OT capacity with $1$-privacy, for honest-but-curious users when the broadcast channel is of the form (\ref{Nan04}), is
\begin{align}
\max_{\gamma_1,\gamma_2,\tau_1,\tau_2}\min\left\{\begin{array}{l}
(\gamma_1-\gamma_2)C_0\\
\gamma_1C_G+\tau_1C_B\\
\gamma_2C_G+\tau_2C_B+(\gamma_1-\gamma_2)C_0\\
\min(\gamma_1+\gamma_2,(1-\epsilon_1))C_G +\min(\tau_1+\tau_2,\epsilon_1)C_B+(\gamma_1-\gamma_2)C_0
\end{array}\right.
\end{align}
where the parameters $\gamma_1,\gamma_2,\tau_1,\tau_2$ satisfy
\begin{align}
\gamma_1+\tau_1=\gamma_2+\tau_2&=\beta\\
0\le\beta&\le1\\
\gamma_1,\gamma_2&\le1-\epsilon_1\\
\tau_1,\tau_2&\le \epsilon_1
\end{align}
\end{Theo}

\begin{IEEEproof}
%
%
In this proof, we will not use the universal$_2$ hash function for compression, but use the technique of the double random binning as in the problem of the secret key generation, suggested by \cite[Proposition 1]{C2009}. We briefly describe our scheme as follows.

We first use a negligible amount of the uses of the noisy channel  to establish  $1$ bit of common randomness $S$ between Alice and Bob, which is secret from Eve.  We will use this random bit to ensure Bob's privacy at Eve.

Then Bob forms the sets $\mathcal{G}$ and $\mathcal{B}$ as follows. We assume that $\mathcal{G}$ consists of $n\gamma_1$ coordinates uniformly selected from the symbols in $\mathcal{Y}_0$ and $n\tau_1$ coordinates from the symbols in $\mathcal{Y}_1$. Similarly, we assume that $\mathcal{B}$ contains $n\gamma_2$ coordinates uniformly selected from the symbols in $\mathcal{Y}_0$ and $n\tau_2$ coordinates from the symbols in $\mathcal{Y}_1$. 
%

The novelty here is that we allow the reuse of coordinates from both $\mathcal{Y}_0$ and $\mathcal{Y}_1$, in contrast to reusing only non-erased coordinates in Theorem $2$. We also allow $\mathcal{G}$ to contain coordinates from both $\mathcal{Y}_0$ and $\mathcal{Y}_1$, while in Theorem 2, only non-erased coordinates is included in $\mathcal{G}$.

We are inspired by the proof of \cite[Theorem 4]{C2009}, more specifically \cite[Proposition 1]{C2009} and \cite[Remark 7]{C2009}, and transform our problem into the secret key generation problem.  We define the following channels,
\begin{align}
W_G(y|x)&=\left\{\begin{array}{cl}W_0(y|x)&\text{ w.p. }\frac{\gamma_1}{\gamma_1+\tau_1}\\W_1(y|x)&\text{ w.p. }\frac{\tau_1}{\gamma_1+\tau_1} \end{array}\right.\\
W_B(y|x)&=\left\{\begin{array}{cl}W_0(y|x)&\text{ w.p. }\frac{\gamma_2}{\gamma_2+\tau_2}\\W_1(y|x)&\text{ w.p. }\frac{\tau_2}{\gamma_2+\tau_2} \end{array}\right.\\
V_G(y|x)&=\left\{\begin{array}{cl}V_0(z|x)&\text{ w.p. }\frac{\gamma_1(1-\epsilon_3)}{\gamma_1+\tau_1}+\frac{\tau_1(1-\epsilon_2)}{\gamma_1+\tau_1}\\V_1(z|x)&\text{ w.p. }\frac{\gamma_1\epsilon_3}{\gamma_1+\tau_1}+\frac{\tau_1\epsilon_2}{\gamma_1+\tau_1} \end{array}\right.\\
V_B(y|x)&=\left\{\begin{array}{cl}V_0(z|x)&\text{ w.p. }\frac{\gamma_2(1-\epsilon_3)}{\gamma_2+\tau_2}+\frac{\tau_2(1-\epsilon_2)}{\gamma_2+\tau_2}\\V_1(z|x)&\text{ w.p. }\frac{\gamma_2\epsilon_3}{\gamma_2+\tau_2}+\frac{\tau_2\epsilon_2}{\gamma_2+\tau_2} \end{array}\right.
\end{align}
We note that $W_G, V_G$ are two branches connecting Alice, Bob and Eve given $X_{\mathcal{G}}$ is sent into the channel, and similarly, $W_B, V_B$ are branches given $X_{\mathcal{B}}$ is sent.
 
We consider the following secret key generation problems.
First, we consider the problem given in \cite[Remark 7]{C2009}, where Alice and  Bob are connected by $W_G$ and the eavesdropper, who is also Bob, is connected to Alice by $W_B$. The resulting rate constraint is 
\begin{align}
R\le \beta(I(P,W_G)-I(P,W_B))=(\gamma_1-\gamma_2)C_0
\end{align}

Secondly, we take Eve into consideration. We assume that Alice and  Bob are connected by $W_G$, and Alice and Eve are connected by $V_G$ while $X_{\mathcal{G}}$ is fed into the channel by Alice. The corresponding rate constraint is
\begin{align}
&R\le \beta(I(P,W_G)-I(P,V_G))
=\gamma_1C_G+\tau_1C_B
\end{align}

Next,  we assume that Alice and  Bob are connected by $W_G$ and Alice and Eve are connected by $V_B$ while $X_{\mathcal{B}}$ is fed into the channel by Alice. From \cite[Proposition 1]{C2009}, we have
\begin{align}
R&\le \beta(I(P,W_G)-I(P,V_B))\nonumber\\
&=\gamma_2C_G+\tau_1C_B+(\gamma_1-\gamma_2)C_N=\gamma_2C_G+\tau_2C_B+(\gamma_1-\gamma_2)C_0
\end{align}

Finally, we consider feeding both $X_{\mathcal{G}}$ and $X_{\mathcal{B}}$ into the channel. Alice and Bob are connected by $W_G$ for both $X_{\mathcal{G}}$ and $X_{\mathcal{B}}$,  and Eve is connected to Alice by $W_G$ for $X_{\mathcal{G}}$ and $W_B$ for $X_{\mathcal{B}}$. Therefore, we have
\begin{align}
2R\le \min(\gamma_1+\gamma_2,n(1-\epsilon_1))C_G +\min(\tau_1+\tau_2,n\epsilon_1)C_B+(\gamma_1-\gamma_2)C_0
\end{align}
which concludes the proof of the theorem. 
\end{IEEEproof}

Let us specify the above theorem to the BESBC. 
\begin{Cor}\label{cor_pro}
For the wiretapped OT problem with BESBC, the following rate is achievable. 
\begin{align}
\left\{ \begin{array}{ll}\min\left(\epsilon_1, (1-\epsilon_1)\epsilon_3,\frac{1}{2}[(1-\epsilon_1)\epsilon_3+\epsilon_1\epsilon_2]\right)&\text{ if }\epsilon_2\ge\epsilon_3\\
\min\left(\epsilon_1, (1-2\epsilon_1)\epsilon_3+\epsilon_1\epsilon_2, \frac{1}{2}[(1-\epsilon_1)\epsilon_3 +\epsilon_1\epsilon_2]\right)&\text{ if }\epsilon_2\le\epsilon_3,\epsilon_1\le\frac{1}{2}\\
(1-\epsilon_1)\epsilon_2&\text{ if }\epsilon_2\le\epsilon_3,\epsilon_1\ge\frac{1}{2}
\end{array}\right.
\end{align}
\end{Cor}
The proof of the corollary can be found in Appendix \ref{pro_cor}.

We note that the achievable rate in the above corollary collapses to the achievable rate in Theorem \ref{Theo:lower bound} if we assume $\epsilon_2\ge\epsilon_3$. 
On the other hand, if $\epsilon_2<\epsilon_3$, the achievable rate in Corollary \ref{cor_pro} does not meet the upper bound in Theorem \ref{Theo:upper bound}. Therefore, the wiretapped OT capacity for the BESBC in the case of $\epsilon_2<\epsilon_3$ is still unknown.

\section{Conclusions}
We proposed a protocol that achieves the 1-of-2 string wiretapped OT capacity for the BESBC when $\epsilon_2 \ge \epsilon_3$. Our result is tighter and more general compared with the state-of-art result of \cite[Theorem 5]{M2017}. 
We  further propose a new  protocol which is applicable to a larger class of broadcast channels and obtain a novel lower bound on the wiretapped OT capacity.

\appendices

\section{Proof of lemma \ref{lem:C1p}} \label{Nan03}
We first list basic definitions and properties regarding universal hash functions, R\'enyi entropy and privacy amplification as follows, which can also be found in \cite[Appendix A]{M2017}.

\begin{Def}
A class $\mathcal{F}$ of functions mapping $\mathcal{A} \longrightarrow \mathcal{B}$ is \emph{universal}$_2$ if, for $F \thicksim \text{Unif}(\mathcal{F})$ and for any $a_0,a_1 \in \mathcal{A}, a_0 \neq a_1$, we have
\begin{align}  P[F(a_0) = F(a_1)] \leq \frac{1}{|\mathcal{B}|}   \end{align}
\end{Def}

The class of all linear maps from $\{0,1\}^n$ to $\{0,1\}^r$ is a universal$_2$ class.

\begin{Def}
Let $A$ be a random variable with alphabet $\mathcal{A}$ and distribution $p_A$. The collision probability $P_c(A)$ of $A$ is defined as the probability that $A$ takes the same value twice in two independent experiments. That is,
\begin{align}  P_c(A) = \underset{a \in \mathcal{A}}{\sum} p^2_A(a) \end{align}
\end{Def}

\begin{Def}
The R\'enyi entropy of order two of a random variable $A$ is
\begin{align}  R(A) = \log_2 \left( \frac{1}{P_c(A)}  \right)  \end{align}
\end{Def}

For an event $\mathcal{E}$, the conditional distribution $p_{A|\mathcal{E}}$ is used to define the conditional collision probability $P_c(A|\mathcal{E})$ and the conditional R\'enyi entropy of order $2$, $R(A|\mathcal{E})$.

\begin{Lem}[Corollary 4 of \cite{B1995}]
\label{lem:privacy_amplification}
Let $P_{AD}$ be an arbitrary probability distribution, with $A \in \mathcal{A}, D \in \mathcal{D}$, and let $d \in \mathcal{D}$. Suppose $R(A | D = d) \geq c$. Let $\mathcal{F}$ be a universal$_2$ class of functions mapping $\mathcal{A} \longrightarrow \{0,1\}^l$ and $F \thicksim \text{Unif}(\mathcal{F})$. Then,
\begin{align}
  H(F(A) | F, D = d) & \geq l -  \log (1 + 2^{l - c}) \\
                             & \geq l - \frac{2^{l-c}}{\ln 2}
\end{align}
\end{Lem}

To proof Lemma~\ref{lem:C1p}, we follow the same proof steps as those in \cite[lemma 9 ]{M2017} and  prove  that (\ref{eqn:ach_1p_wtap_0}) - (\ref{eqn:ach_1p_wtap_3}) are satisfied for our protocol.

Suppose that when the length of the erased coordinates received by Bob is less than $n(\epsilon_1-\delta)$ or the length of the non-erased coordinates is less than $n(1-\epsilon_1-\delta)$, Bob will abort the protocol and request Alice to resend $X^n$ to the noisy channel. 
We use $\#(e(Y^n))$ and $\#(\bar{e}(Y^n))$ to represent the number of erased and non-erased coordinates of $Y^n$, respectively. From \cite[Lemma 2.6]{csiszar2011}, we have that the probabilities 
\begin{align}
P[\#(e(Y^n)) \geq n(\epsilon_1-\delta)]&\geq 1-K(n)\exp(-nD(\epsilon_1-\delta||\epsilon_1))\label{P1}\\
P[\#(\bar{E}(Y^n)) \geq n(1-\epsilon_1-\delta)]&\geq 1-K(n)\exp(-nD(1-\epsilon_1-\delta||1-\epsilon_1))\label{P2}
\end{align}
where $K(n)$ is a polynomial of  $n$, and for $0\le p,q\le1$
\begin{align}
D(p||q)\triangleq p\log\frac{p}{q}+(1-p)\log\frac{1-p}{1-q}
\end{align}
We can see that with sufficiently large $n$, the probabilities in (\ref{P1}) and (\ref{P2}) are arbitrarily close to $1$.

\emph{Correctness}:
We prove that (\ref{eqn:ach_1p_wtap_0}) is satisfied for our protocol. Since Bob knows $X_{\mathcal{L}_\alpha}$, he can compute the key $F_\alpha(X_{\mathcal{L}_\alpha})$ and obtain its first bit $K_\alpha$. From $K_\alpha$ computed by Bob and $K_\alpha\oplus S$ transmitted by Alice through public channel, Bob can retrieve the value of $S$. Similarly, Bob knows $X_{\mathcal{G}}=X_{\mathcal{L}_{\Theta\oplus S}}$ and $F_{\Theta\oplus S}$, then Bob can compute the key $F_{\Theta\oplus S}(X_{\mathcal{L}_{\Theta\oplus S}})$. As a result, Bob can learn the value of $K_\Theta$ from $K_\Theta\oplus F_{\Theta\oplus S}(X_{\mathcal{L}_{\Theta\oplus S}})$ sent by Alice and the key $F_{\Theta\oplus S}(X_{\mathcal{L}_{\Theta\oplus S}})$. Hence, $P [ \hat{K}_\Theta \neq K_\Theta ]=0$.           

\emph{Security at Bob}:
We prove that (\ref{eqn:ach_1p_wtap_1}) is satisfied for our protocol.
\begin{align}
    I(K_{\bar{\Theta}}; V_B) &= I(K_{\bar{\Theta}}; \Theta, Y, N, F)\nonumber\\
    &=I(K_{\bar{\Theta}}; \Theta, Y, N ,\mathcal{L}_0, \mathcal{L}_1, \mathcal{L}_a, S\oplus K_a, \nonumber\\& \quad \quad F_\alpha, F_0, F_1, K_{\Theta} \oplus F_{\Theta\oplus S}(X_{\mathcal{L}_{\Theta\oplus S}}), K_{\bar{\Theta}} \oplus  F_{\bar{\Theta}\oplus S}(X_{\mathcal{L}_{\bar{\Theta}\oplus S}}))\nonumber\\
    &\overset{(a)}{=}I(K_{\bar{\Theta}}; \Theta, Y^n, N, \mathcal{L}_{\Theta}, \mathcal{L}_{\bar{\Theta}}, \mathcal{L}_\alpha, S, \nonumber\\& \quad \quad F_\alpha, F_{\Theta}, F_{\bar{\Theta}}, K_{\Theta} \oplus F_{\Theta\oplus S}(X_{\mathcal{L}_{\Theta\oplus S}}), K_{\bar{\Theta}} \oplus  F_{\bar{\Theta}\oplus S}(X_{\mathcal{L}_{\bar{\Theta}\oplus S}}))\nonumber\\
    &\overset{(b)}{=}I(K_{\bar{\Theta}}; \Theta, Y^n, N, \mathcal{L}_{\Theta}, \mathcal{L}_{\bar{\Theta}}, \mathcal{L}_a, S, F_\alpha, F_{\Theta}, F_{\bar{\Theta}}, K_{\Theta}, K_{\bar{\Theta}} \oplus  F_{\bar{\Theta}\oplus S}(X_{\mathcal{L}_{\bar{\Theta}\oplus S}}))\nonumber\\
    &\overset{(c)}{=}I(K_{\bar{\Theta}}; \Theta, Y^n, N, \mathcal{L}_{\Theta}, \mathcal{L}_{\bar{\Theta}}, \mathcal{L}_a, S, F_\alpha, F_{\Theta}, F_{\bar{\Theta}}, K_{\bar{\Theta}} \oplus  F_{\bar{\Theta}\oplus S}(X_{\mathcal{L}_{\bar{\Theta}\oplus S}}))\nonumber\\
    &\overset{(d)}{=}I(K_{\bar{\Theta}}; K_{\bar{\Theta}} \oplus  F_{\bar{\Theta}\oplus S}(X_{\mathcal{L}_{\bar{\Theta}\oplus S}})| \Theta, Y^n, N, \mathcal{L}_{\Theta}, \mathcal{L}_{\bar{\Theta}}, \mathcal{L}_\alpha, S, F_\alpha, F_{\Theta}, F_{\bar{\Theta}})\nonumber\\
    &=H(K_{\bar{\Theta}} \oplus  F_{\bar{\Theta}\oplus S}(X_{\mathcal{L}_{\bar{\Theta}\oplus S}})| \Theta, Y^n, N, \mathcal{L}_{\Theta}, \mathcal{L}_{\bar{\Theta}}, \mathcal{L}_\alpha, S, F_\alpha, F_{\Theta}, F_{\bar{\Theta}})\nonumber\\
    &\quad -H(F_{\bar{\Theta}\oplus S}(X_{\mathcal{L}_{\bar{\Theta}\oplus S}})|K_{\bar{\Theta}}, \Theta, Y^n, N, \mathcal{L}_{\Theta}, \mathcal{L}_{\bar{\Theta}}, \mathcal{L}_\alpha, S, F_\alpha, F_{\Theta}, F_{\bar{\Theta}})\nonumber\\
    &\leq \log|F_{\bar{\Theta}\oplus S}(X_{\mathcal{L}_{\bar{\Theta}\oplus S}})| -H(F_{\bar{\Theta}\oplus S}(X_{\mathcal{L}_{\bar{\Theta}\oplus S}})|K_{\bar{\Theta}}, \Theta, Y^n, N, \mathcal{L}_{\Theta}, \mathcal{L}_{\bar{\Theta}}, \mathcal{L}_\alpha, S, F_\alpha, F_{\Theta}, F_{\bar{\Theta}})\nonumber\\
    &\overset{(e)}{=} n(R-\tilde{\delta})-H(F_{\bar{\Theta}\oplus S}(X_{\mathcal{L}_{\bar{\Theta}\oplus S}})|K_{\bar{\Theta}}, \Theta, Y^n, N, \mathcal{L}_{\Theta}, \mathcal{L}_{\bar{\Theta}}, \mathcal{L}_\alpha, S, F_\alpha, F_{\Theta}, F_{\bar{\Theta}})\nonumber\\
    &\overset{(f)}{=}  n(R-\tilde{\delta})-H(F_{\bar{\Theta}\oplus S}(X_{\mathcal{L}_{\bar{\Theta}\oplus S}})| Y_{\mathcal{L}_{\bar{\Theta}\oplus S}},F_{\bar{\Theta}\oplus S})
\end{align}
where 
\begin{enumerate}[label=(\alph*)]
\item holds since $K_\alpha$ is a function of $(F_\alpha, Y^n, \mathcal{L}_\alpha)$; 
\item  holds since $F_{\Theta\oplus S}(X_{\mathcal{L}_{\Theta\oplus S }})$ is a function of $(F_{\Theta\oplus S}, Y^n, \mathcal{L}_{\Theta\oplus S})$; 
\item  holds since $K_\Theta$ is independent of all other  random variables in the mutual information term; 
\item  holds since $K_{\bar{\Theta}}$ is independent of $(\Theta, Y^n, N, \mathcal{L}_{\Theta}, \mathcal{L}_{\bar{\Theta}}, \mathcal{L}_\alpha, S, F_\alpha, F_{\Theta}, F_{\bar{\Theta}})$; 
\item  holds since the length of the sequence $F_{\bar{\Theta}}(X_{\mathcal{L}_{\bar{\Theta}}})$ is $n(R-\tilde \delta)$; 
\item  holds because of the following Markov chain
\begin{align}
F_{\bar{\Theta}\oplus S}(X_{\mathcal{L}_{\bar{\Theta}\oplus S}})\rightarrow(Y_{\mathcal{L}_{\bar{\Theta}\oplus S}},F_{\bar{\Theta}\oplus S})\rightarrow (K_{\bar{\Theta}}, \Theta, Y^n, N, \mathcal{L}_{\Theta}, \mathcal{L}_{\bar{\Theta}}, \mathcal{L}_\alpha, S, F_\alpha, F_{\Theta}, F_{\bar{\Theta}})
\end{align} 
\end{enumerate}
We note that $\mathcal{L}_{\bar{\Theta}\oplus S}=\mathcal{B}$
\begin{align}
    R(X_{\mathcal{L}_{\mathcal{B}}}|Y_{\mathcal{L}_{\mathcal{B}}}=y_{\mathcal{L}_{\mathcal{B}}}) &=\#(e(y_{\mathcal{L}_{\mathcal{B}}}))
\end{align}
 where $\#(e(y_{\mathcal{L}_{\mathcal{B}}}))$ represents the number of coordinates, at which $y_{\mathcal{L}_{\mathcal{B}}}$ are erased. We note that
 \begin{align}
 \mathsf{Pr}[\#(e(Y_{\mathcal{L}_{\mathcal{B}}})) \geq n(R-\delta)]
 \ge\mathsf{Pr}[\#(e(Y^n)) \geq n(\epsilon_1-\delta)]
 &\geq 1-K(n)\exp(-nD(\epsilon_1-\delta||\epsilon_1))
 \end{align}
 
By defining $\xi_1\triangleq K(n)\exp(-nD(\epsilon_1-\delta||\epsilon_1))$ and using Lemma~\ref{lem:privacy_amplification}, we have
\begin{align}
    &H(F_{\bar{\Theta}\oplus S}(X_{\mathcal{L}_{\bar{\Theta}\oplus S}})| Y_{\mathcal{L}_{\bar{\Theta}\oplus S}},F_{\bar{\Theta}\oplus S})\nonumber\\ &\geq 
 \mathsf{Pr}[\#(e(Y_{\mathcal{L}_{\mathcal{B}}})) \geq n(R-\delta)] H(F_{\bar{\Theta}\oplus S}(X_{\mathcal{L}_{\bar{\Theta}\oplus S}})| Y_{\mathcal{L}_{\bar{\Theta}\oplus S}},F_{\bar{\Theta}\oplus S},\#_E(Y_{\mathcal{L}_{\mathcal{B}}}) \geq n(R-\delta))   \nonumber\\ &\geq
  (1-\xi_1)  \left(n(R-\tilde \delta) - \frac{2^{n(R-\tilde \delta)-n(R-\delta)}}{\ln2}\right)\nonumber\\ 
    &=  (1-\xi_1)  \left(n(R-\tilde \delta) - \frac{2^{-n(\tilde \delta -\delta)}}{\ln2} \right)
\end{align}
Thus,
\begin{align}
     I(K_{\bar{\Theta}}; V_B) &\leq  n(R-\tilde{\delta})-H(F_{\bar{\Theta}\oplus S}(X_{\mathcal{L}_{\bar{\Theta}\oplus S}})| Y_{\mathcal{L}_{\bar{\Theta}\oplus S}},F_{\bar{\Theta}\oplus S})\nonumber\\
     &\leq n(R-\tilde{\delta})-(1-\xi_1)  \left(n(R-\tilde \delta) - \frac{2^{-n(\tilde \delta -\delta)}}{\ln2} \right)\nonumber\\
     &=\xi_1n(R-\tilde{\delta})+(1-\xi_1)\frac{2^{-n(\tilde \delta -\delta)}}{\ln2}
\end{align}
Therefore, $I(K_{\bar{\Theta}}; V_B) \to 0$ as $n \to \infty$ if we assume $\delta<\bar\delta$.

\emph{Privacy at Alice}:
We prove that (\ref{eqn:ach_1p_wtap_2}) is satisfied for our protocol. 
\begin{align}
    &I(\Theta;V_A)=I(\Theta; K_0, K_1, X^n, M, F)\nonumber\\
    &\overset{(a)}{=}I(\Theta; K_0, K_1, X^n, M, S, F)\nonumber\\
    &=I(\Theta; K_0, K_1, X^n, M, S, \mathcal{L}_0, \mathcal{L}_1, \mathcal{L}_\alpha, S\oplus K_\alpha, F_\alpha, F_0, F_1, K_{\Theta} \oplus F_{\Theta\oplus S}(X_{\mathcal{L}_{\Theta\oplus S}}), K_{\bar{\Theta}} \oplus  F_{\bar{\Theta}\oplus S}(X_{\mathcal{L}_{\bar{\Theta}\oplus S}}))\nonumber\\
    &=I(\Theta; K_0, K_1, X^n, M, S, \mathcal{L}_0, \mathcal{L}_1, \mathcal{L}_\alpha,  K_\alpha, F_\alpha, F_0, F_1,  K_{\Theta} \oplus F_{\Theta\oplus S}(X_{\mathcal{L}_{\Theta\oplus S}}), K_{\bar{\Theta}} \oplus  F_{\bar{\Theta}\oplus S}(X_{\mathcal{L}_{\bar{\Theta}\oplus S}}))\nonumber\\
    &\overset{(b)}{=}I(\Theta; S, \mathcal{L}_0, \mathcal{L}_1)\nonumber\\
    &\overset{(c)}{=}I(\Theta; \mathcal{L}_0, \mathcal{L}_1|S)\nonumber\\
    &\overset{(d)}{=}0
\end{align}
where 
\begin{enumerate}[label=(\alph*)]
\item follows since $S$ is a function of $M$; 
\item follows because of the following Markov chain
\begin{align}
\Theta\rightarrow(S, \mathcal{L}_0, \mathcal{L}_1)\rightarrow (K_0,K_1,X^n, M,\mathcal{L}_\alpha,  K_\alpha, F_\alpha, F_0, F_1, K_{\Theta} \oplus F_{\Theta\oplus S}(X_{\mathcal{L}_{\Theta\oplus S}}), K_{\bar{\Theta}} \oplus  F_{\bar{\Theta}\oplus S}(X_{\mathcal{L}_{\bar{\Theta}\oplus S}}))
\end{align}
\item follows since $\Theta$ and $S$ are independent; 
\item follows since given  $ S,\Theta$, the distribution of $(\mathcal{L}_0, \mathcal{L}_1)$ is uniform in $\{1,2\dots,n\}$, which is same as the distribution of $(\mathcal{L}_0, \mathcal{L}_1)$ given $S$.
\end{enumerate}

\emph{Security and Privacy at Eve}:
We prove that (\ref{eqn:ach_1p_wtap_3}) is satisfied for out protocol. 
\begin{align}
    I\left(K_{0}, K_{1}, \Theta ; V_{E}\right) &=I\left(K_{\Theta}, K_{\bar{\Theta}}, \Theta ; V_{E}\right) \nonumber\\
&=I\left(\Theta ; V_{E}\right)+I\left(K_{\bar{\Theta}} ; V_{E} \mid \Theta\right)+I\left(K_{\Theta} ; V_{E} \mid \Theta, K_{\bar{\Theta}}\right) \nonumber\\
&=I\left(\Theta ; V_{E}\right)+I\left(K_{\bar{\Theta}} ; \Theta, V_{E}\right)+I\left(K_{\Theta} ; \Theta, K_{\bar{\Theta}}, V_{E}\right)
\end{align}
For the first term, we have
\begin{align}
  &  I(\Theta; V_{E})=I(\Theta; Z^n, F)\nonumber\\
&=I(\Theta; Z^n, \mathcal{L}_{\alpha},S \oplus K_\alpha, F_{\alpha}, \mathcal{L}_0, \mathcal{L}_1, F_0, F_1, K_{\Theta} \oplus F_{\Theta\oplus S}(X_{\mathcal{L}_{\Theta\oplus S}}),K_{\bar{\Theta}} \oplus F_{\bar{\Theta}\oplus S}(X_{\mathcal{L}_{\bar{\Theta}\oplus S}}))\nonumber\\
&\leq I(\Theta; Z^n, \mathcal{L}_{\alpha},S \oplus K_\alpha, F_{\alpha}, \mathcal{L}_0, \mathcal{L}_1, F_0, F_1,K_{\Theta},  K_{\Theta} \oplus F_{\Theta\oplus S}(X_{\mathcal{L}_{\Theta\oplus S}}),K_{\bar{\Theta}}, K_{\bar{\Theta}} \oplus F_{\bar{\Theta}\oplus S}(X_{\mathcal{L}_{\bar{\Theta}\oplus S}}))\nonumber\\
&=I(\Theta;Z^n,\mathcal{L}_{a},S \oplus K_\alpha, F_{\alpha},\mathcal{L}_0,\mathcal{L}_1,F_0,F_1 ,K_{\Theta}, F_{\Theta \oplus S}(X_{\mathcal{L}_{\Theta \oplus S}}),K_{\bar{\Theta}},  F_{\bar{\Theta}\oplus S }(X_{\mathcal{L}_{\bar{\Theta}\oplus S }}))\nonumber\\
&\overset{(a)}{=}I(\Theta; Z^n, \Theta\oplus S, \mathcal{L}_{a}, S \oplus K_\alpha, F_{a},\mathcal{L}_0,\mathcal{L}_1,F_0,F_1 ,F_{\Theta \oplus S}(X_{\mathcal{L}_{\Theta \oplus S}}), F_{\bar{\Theta}\oplus S }(X_{\mathcal{L}_{\bar{\Theta}\oplus S }}))\nonumber\\
&=I(\Theta; Z^n,\mathcal{L}_{\alpha},S \oplus K_\alpha, \mathcal{L}_0, \mathcal{L}_1,F_0,F_1 ,F_{\Theta \oplus S}(X_{\mathcal{L}_{\Theta \oplus S}}), F_{\bar{\Theta}\oplus S }(X_{\mathcal{L}_{\bar{\Theta}\oplus S }})|\Theta\oplus S)\nonumber\\ &\quad+I(\Theta;\Theta\oplus S)\nonumber\\ 
&\overset{(b)}{=}I(\Theta\oplus \Theta \oplus S; Z^n, \mathcal{L}_{\alpha}, S \oplus K_\alpha, F_{\alpha}, \mathcal{L}_0,\mathcal{L}_1,F_0,F_1 ,F_{\Theta \oplus S}(X_{\mathcal{L}_{\Theta \oplus S}}), F_{\bar{\Theta}\oplus S }(X_{\mathcal{L}_{\bar{\Theta}\oplus S }})|\Theta\oplus S)\nonumber\\
&=I(S;Z^n, \mathcal{L}_{\alpha},S \oplus K_\alpha, F_{\alpha},\mathcal{L}_0,\mathcal{L}_1,F_0,F_1 ,F_{\Theta \oplus S}(X_{\mathcal{L}_{\Theta \oplus S}}), F_{\bar{\Theta}\oplus S }(X_{\mathcal{L}_{\bar{\Theta}\oplus S }})|\Theta\oplus S)\nonumber\\
&=I(S;S \oplus K_\alpha |Z^n, \mathcal{L}_{\alpha}, F_{\alpha},\mathcal{L}_0,\mathcal{L}_1,F_0,F_1 ,F_{\Theta \oplus S}(X_{\mathcal{L}_{\Theta \oplus S}}), F_{\bar{\Theta}\oplus S }(X_{\mathcal{L}_{\bar{\Theta}\oplus S }}), \Theta\oplus S)\nonumber\\
&=H(S \oplus K_\alpha |Z^n, \mathcal{L}_{\alpha},F_{\alpha},\mathcal{L}_0,\mathcal{L}_1,F_0,F_1 ,F_{\Theta \oplus S}(X_{\mathcal{L}_{\Theta \oplus S}}), F_{\bar{\Theta}\oplus S }(X_{\mathcal{L}_{\bar{\Theta}\oplus S }}),\Theta\oplus S)\nonumber\\
&\quad-H(S \oplus K_\alpha|S,Z^n,\mathcal{L}_{\alpha},F_{\alpha},\mathcal{L}_0,\mathcal{L}_1,F_0,F_1 ,F_{\Theta \oplus S}(X_{\mathcal{L}_{\Theta \oplus S}}), F_{\bar{\Theta}\oplus S }(X_{\mathcal{L}_{\bar{\Theta}\oplus S }}),\Theta\oplus S)\nonumber\\
&\leq |S \oplus K_\alpha| 
- H(K_\alpha|S,Z^n,\mathcal{L}_{\alpha},F_{\alpha},\mathcal{L}_0,\mathcal{L}_1,F_0,F_1 ,F_{\Theta \oplus S}(X_{\mathcal{L}_{\Theta \oplus S}}), F_{\bar{\Theta}\oplus S }(X_{\mathcal{L}_{\bar{\Theta}\oplus S }}),\Theta\oplus S)\nonumber\\
&\leq |S \oplus K_\alpha| - H(K_\alpha|S,Z^n,\mathcal{L}_{\alpha},F_{\alpha},\mathcal{L}_0,\mathcal{L}_1,F_0,F_1 ,F_{\Theta \oplus S}(X_{\mathcal{L}_{\Theta \oplus S}}), F_{\bar{\Theta}\oplus S }(X_{\mathcal{L}_{\bar{\Theta}\oplus S }}),\Theta\oplus S)\nonumber\\ &\quad - H(F_{\alpha}(X_{\mathcal{L}_{\alpha}})|K_\alpha,S,Z^n,\mathcal{L}_{\alpha},F_{\alpha},\mathcal{L}_0,\mathcal{L}_1,F_0,F_1 ,F_{\Theta \oplus S}(X_{\mathcal{L}_{\Theta \oplus S}}), F_{\bar{\Theta}\oplus S }(X_{\mathcal{L}_{\bar{\Theta}\oplus S }}),\Theta\oplus S)\nonumber\\ & \quad +|F_{\alpha}(X_{\mathcal{L}_{\alpha}})|-1\nonumber\\
&=|F_{\alpha}(X_{\mathcal{L}_{\alpha}})| - H(F_{\alpha}(X_{\mathcal{L}_{\alpha}}), K_\alpha|S,Z^n,\mathcal{L}_{\alpha},F_{\alpha},\mathcal{L}_0,\mathcal{L}_1,F_0,F_1 ,F_{\Theta \oplus S}(X_{\mathcal{L}_{\Theta \oplus S}}), F_{\bar{\Theta}\oplus S }(X_{\mathcal{L}_{\bar{\Theta}\oplus S }}),\Theta\oplus S)\nonumber\\
&\overset{(c)}{=}|F_{\alpha}(X_{\mathcal{L}_{\alpha}})| - H(F_{\alpha}(X_{\mathcal{L}_{\alpha}})|S,Z^n,\mathcal{L}_{\alpha},F_{\alpha},\mathcal{L}_0,\mathcal{L}_1,F_0,F_1 ,F_{\Theta \oplus S}(X_{\mathcal{L}_{\Theta \oplus S}}), F_{\bar{\Theta}\oplus S }(X_{\mathcal{L}_{\bar{\Theta}\oplus S }}),\Theta\oplus S)\nonumber\\
&\overset{(d)}{=}|F_{\alpha}( X_{\mathcal{L}_{\alpha}})| - H(F_{\alpha}( X_{\mathcal{L}_{\alpha}})|Z_{\mathcal{L}_{\alpha}},F_{\alpha})\nonumber\\
&\overset{(e)}{\leq} n(\alpha (\epsilon_3- \delta) - \bar \delta)-(1-\xi_3)\left( n(\alpha (\epsilon_3- \delta)-  \bar \delta) -  \frac{2^{-n\delta}}{\ln2}\right)
\end{align}
where 
\begin{enumerate}[label=(\alph*)]
\item follows since $\Theta \oplus S $ is a function of $(Z^n,\mathcal{L}_0,\mathcal{L}_1)$, and $(K_{\Theta},K_{\bar{\Theta}})$ is independent of all the other random variables in the mutual information term; 
\item follows since $\Theta$ is independent of $\Theta \oplus S$; 
\item follows since $K_\alpha$ is a function of $F_{\alpha}( X_{\mathcal{L}_{\alpha}})$; 
\item follows because of the Markov chain
\begin{align}
F_{\alpha}( X_{\mathcal{L}_{\alpha}})\rightarrow (Z_{L_{\alpha}},F_{\alpha}) \rightarrow (S,Z^n,\mathcal{L}_\alpha,\mathcal{L}_0,\mathcal{L}_1,F_0,F_1 ,F_{\Theta \oplus S}(X_{\mathcal{L}_{\Theta \oplus S}}), F_{\bar{\Theta}\oplus S }(X_{\mathcal{L}_{\bar{\Theta}\oplus S }}),\Theta\oplus S)
\end{align}
\item follows because $R(X_{\mathcal{L}_{a}} | Z_{\mathcal{L}_\alpha}=z_{\mathcal{L}_\alpha}) = \#(e(z_{\mathcal{L}_\alpha}))$
\begin{align}
&H(F_{\alpha}( X_{\mathcal{L}_{\alpha}})|Z_{\mathcal{L}_{\alpha}},F_{\alpha})\nonumber\\
&\ge \mathsf{Pr}(\#(e(Z_{\mathcal{L}_\alpha}))\ge(\epsilon_3 -\delta)|\mathcal{L}_\alpha|)H(F_{\alpha}( X_{\mathcal{L}_{\alpha}})|Z_{\mathcal{L}_{\alpha}},F_{\alpha}, \#(e(Z_{\mathcal{L}_\alpha}))\ge(\epsilon_3 -\delta)|\mathcal{L}_\alpha|)
\end{align}
We note that
\begin{align}
P[\#(e(Z_{\mathcal{L}_\alpha})) \geq (\epsilon_3 -\delta)|\mathcal{L}_\alpha|] &\geq 1-\xi_3\\
\xi_3&\triangleq K(n)\exp\left(-n\alpha D(\epsilon_3||\epsilon_3-\delta)\right)
\end{align} 
We note that if $R(X_{\mathcal{L}_{a}} | Z_{\mathcal{L}_\alpha}) = \#(e(Z_{\mathcal{L}_\alpha})) \geq n\alpha(\epsilon_3- \delta)$, we have 
\begin{align}
H(F_{\alpha}( X_{\mathcal{L}_{\alpha}})|&Z_{\mathcal{L}_{\alpha}},F_{\alpha}, \#(e(Z_{\mathcal{L}_\alpha}))\ge(\epsilon_3 -\delta)|\mathcal{L}_\alpha|) \nonumber\\
&\geq  n(\alpha (\epsilon_3- \delta)-  \bar \delta) -  \frac{2^{-n\delta}}{\ln2} 
\end{align}
Then, 
\begin{align}
H(F_{\alpha}( X_{\mathcal{L}_{\alpha}})|Z_{\mathcal{L}_{\alpha}},F_{\alpha})\ge(1-\xi_3)\left(  n(\alpha (\epsilon_3- \delta)-  \bar \delta) -  \frac{2^{-n\delta}}{\ln2}\right)
\end{align}
\end{enumerate}
From above derivation, we have
\begin{align}
I(\Theta; V_{E})\le\xi_3 n (\alpha (\epsilon_3- \delta) - \bar \delta) + (1-\xi_3) \frac{2^{-n\bar{\delta}}}{\ln2}
\end{align}
which goes to zero with $n\to \infty$.

For the second term, we have
\begin{align}
    &I(K_{\bar{\Theta}};\Theta,V_E)=I(K_{\bar{\Theta}};\Theta,Z^n,F)\nonumber\\
    &=I(K_{\bar{\Theta}};\Theta, Z^n, \mathcal{L}_0, \mathcal{L}_1, \mathcal{L}_\alpha, S\oplus K_\alpha, F_\alpha, F_0, F_1, K_{\Theta} \oplus F_{\Theta\oplus S}(X_{L_{\Theta\oplus S}}), \nonumber\\&\quad \quad K_{\bar{\Theta}} \oplus  F_{\bar{\Theta}\oplus S}(X_{\mathcal{L}_{\bar{\Theta}\oplus S}}))\nonumber\\
    &\overset{(a)}{=}I(K_{\bar{\Theta}};\Theta,Z^n,S, \mathcal{L}_{\Theta\oplus S}, \mathcal{L}_{\bar{\Theta}\oplus S}, \mathcal{L}_\alpha, S\oplus K_\alpha, F_\alpha, F_{\Theta\oplus S}, F_{\bar{\Theta}\oplus S}, K_{\Theta} \oplus F_{\Theta\oplus S}(X_{\mathcal{L}_{\Theta\oplus S}}), \nonumber\\&\quad \quad K_{\bar{\Theta}} \oplus  F_{\bar{\Theta}\oplus S}(X_{\mathcal{L}_{\bar{\Theta}\oplus S}}))\nonumber\\
    &=I(K_{\bar{\Theta}};\Theta,Z^n,S, \mathcal{L}_{\Theta\oplus S}, \mathcal{L}_{\bar{\Theta}\oplus S}, \mathcal{L}_\alpha, K_\alpha, F_\alpha, F_{\Theta\oplus S}, F_{\bar{\Theta}\oplus S}, K_{\Theta} \oplus F_{\Theta\oplus S}(X_{\mathcal{L}_{\Theta\oplus S}}), \nonumber\\&\quad \quad K_{\bar{\Theta}} \oplus  F_{\bar{\Theta}\oplus S}(X_{\mathcal{L}_{\bar{\Theta}\oplus S}}))\nonumber\\
    &\overset{(b)}{\le} I(K_{\bar{\Theta}};\Theta,Z^n,S, \mathcal{L}_{\Theta\oplus S}, \mathcal{L}_{\bar{\Theta}\oplus S}, \mathcal{L}_\alpha, K_\alpha, F_\alpha, F_{\Theta\oplus S}, F_{\bar{\Theta}\oplus S}, F_{\Theta\oplus S}(X_{\mathcal{L}_{\Theta\oplus S}}), \nonumber\\&\quad \quad K_{\bar{\Theta}} \oplus  F_{\bar{\Theta}\oplus S}(X_{\mathcal{L}_{\bar{\Theta}\oplus S}}))\nonumber\\
    &=I(K_{\bar{\Theta}};K_{\bar{\Theta}} \oplus  F_{\bar{\Theta}\oplus S}(X_{\mathcal{L}_{\bar{\Theta}\oplus S}}) | \Theta,Z^n,S, \mathcal{L}_{\Theta\oplus S}, \mathcal{L}_{\bar{\Theta}\oplus S}, \mathcal{L}_\alpha, K_\alpha, F_\alpha, F_{\Theta\oplus S}, F_{\bar{\Theta}\oplus S},\nonumber\\&\quad \quad  F_{\Theta\oplus S}(X_{\mathcal{L}_{\Theta\oplus S}}))\nonumber\\
    &=H(K_{\bar{\Theta}} \oplus  F_{\bar{\Theta}\oplus S}(X_{\mathcal{L}_{\bar{\Theta}\oplus S}}) |\Theta,Z^n,S, \mathcal{L}_{\Theta\oplus S}, \mathcal{L}_{\bar{\Theta}\oplus S}, \mathcal{L}_\alpha, K_\alpha, F_\alpha, F_{\Theta\oplus S}, F_{\bar{\Theta}\oplus S},\nonumber\\&\quad \quad  F_{\Theta\oplus S}(X_{\mathcal{L}_{\Theta\oplus S}}))\nonumber\\& \quad - H(K_{\bar{\Theta}} \oplus  F_{\bar{\Theta}\oplus S}(X_{\mathcal{L}_{\bar{\Theta}\oplus S}}) | K_{\bar{\Theta}},\Theta,Z^n,S, \mathcal{L}_{\Theta\oplus S}, \mathcal{L}_{\bar{\Theta}\oplus S}, \mathcal{L}_\alpha, K_\alpha, F_\alpha, F_{\Theta\oplus S}, F_{\bar{\Theta}\oplus S},\nonumber\\&\quad \quad  F_{\Theta\oplus S}(X_{\mathcal{L}_{\Theta\oplus S}}))\nonumber\\
    &\leq |K_{\bar{\Theta}} \oplus  F_{\bar{\Theta}\oplus S}(X_{\mathcal{L}_{\bar{\Theta}\oplus S}})| - H( F_{\bar{\Theta}\oplus S}(X_{\mathcal{L}_{\bar{\Theta}\oplus S}}) | K_{\bar{\Theta}},\Theta,Z^n,S, \mathcal{L}_{\Theta\oplus S}, \mathcal{L}_{\bar{\Theta}\oplus S}, \mathcal{L}_\alpha, \nonumber\\&\quad \quad  K_\alpha, F_\alpha, F_{\Theta\oplus S}, F_{\bar{\Theta}\oplus S}, F_{\Theta\oplus S}(X_{\mathcal{L}_{\Theta\oplus S}}))\nonumber\\
    &\leq |K_{\bar{\Theta}} \oplus  F_{\bar{\Theta}\oplus S}(X_{\mathcal{L}_{\bar{\Theta}\oplus S}})| - H( F_{\bar{\Theta}\oplus S}(X_{\mathcal{L}_{\bar{\Theta}\oplus S}}) | K_{\bar{\Theta}},\Theta, Z^n, S, \mathcal{L}_{\Theta\oplus S}, \mathcal{L}_{\bar{\Theta}\oplus S}, \mathcal{L}_\alpha, \nonumber\\&\quad \quad K_\alpha, F_\alpha, F_{\Theta\oplus S}, F_{\bar{\Theta}\oplus S}, X_{\mathcal{L}_{\Theta\oplus S}})\nonumber\\
    &\overset{(c)}{=} |K_{\bar{\Theta}} \oplus  F_{\bar{\Theta}\oplus S}(X_{\mathcal{L}_{\bar{\Theta}\oplus S}})| - H( F_{\bar{\Theta}\oplus S}(X_{\mathcal{L}_{\bar{\Theta}\oplus S}}) | X_{\mathcal{L}_{b}},\mathcal{L}_{b} , Z _{\mathcal{L}_{\bar{\Theta}\oplus S}}, F_{\bar{\Theta}\oplus S})\nonumber\\
    &\overset{(d)}{\leq} n(R-\tilde \delta)-(1-\xi_5)\left(  n(R-\tilde \delta) -\frac{2^{n(R-\tilde \delta)-n(R-\delta)}}{\ln2} \right)\nonumber\\ 
    &=\xi_5(n(R-\tilde \delta))+(1-\xi_5)\frac{2^{-n(\tilde \delta -\delta)}}{\ln2}
\end{align}
where 
\begin{enumerate}[label=(\alph*)]
\item holds since $S$ is a function of $(\Theta, Z^n, \mathcal{L}_0,\mathcal{L}_1)$;
\item holds because of the Markov chain
\begin{align}
&K_{\Theta} \oplus F_{\Theta\oplus S}(X_{\mathcal{L}_{\Theta\oplus S}})\rightarrow F_{\Theta\oplus S}(X_{\mathcal{L}_{\Theta\oplus S}})\rightarrow\nonumber\\
&  (K_{\bar{\Theta}};\Theta,Z^n,S, \mathcal{L}_{\Theta\oplus S}, \mathcal{L}_{\bar{\Theta}\oplus S}, \mathcal{L}_\alpha, K_\alpha, F_\alpha, F_{\Theta\oplus S}, F_{\bar{\Theta}\oplus S},  K_{\bar{\Theta}} \oplus  F_{\bar{\Theta}\oplus S}(X_{\mathcal{L}_{\bar{\Theta}\oplus S}}))
\end{align}
\item holds because of the Markov chain 
\begin{align}
F_{\bar{\Theta}\oplus S}(X_{\mathcal{L}_{\bar{\Theta}\oplus S}}) \rightarrow(Z _{\mathcal{L}_{\bar{\Theta}\oplus S}}, F_{\bar{\Theta}\oplus S}, X_{\mathcal{L}_{b}},\mathcal{L}_{b})  \rightarrow (K_{\bar{\Theta}},\Theta, Z^n, S, X_{\mathcal{L}_{\Theta\oplus S}},\mathcal{L}_{\Theta\oplus S}, \mathcal{L}_{\bar{\Theta}\oplus S}, \mathcal{L}_\alpha, K_\alpha, F_\alpha, F_{\Theta\oplus S})
\end{align} 
where $\mathcal{L}_b =  \mathcal{L}_{\Theta\oplus S} \cap \mathcal{L}_{\bar{\Theta}\oplus S } $ is the overlapping part between $\mathcal{L}_0$ and $\mathcal{L}_1$;
\item holds because 
\begin{align}
R&(X_{\mathcal{L}_{\bar{\Theta} \oplus S}} | X_{\mathcal{L}_b}, \mathcal{L}_b, Z_{\mathcal{L}_{\bar{\Theta} \oplus S}}=z_{\mathcal{L}_{\bar{\Theta} \oplus S}}) 
=\#(e(z_{\mathcal{L}_{\bar{\Theta}\oplus S}\backslash{\mathcal{L}_b}}))\nonumber\\
&=\#(e(y_{\mathcal{L}_{\bar{\Theta}\oplus S}\backslash{\mathcal{L}_b}})\wedge e(z_{\mathcal{L}_{\bar{\Theta}\oplus S}\backslash{\mathcal{L}_b}}))+
\#(\bar{e}(y_{\mathcal{L}_{\bar{\Theta}\oplus S}\backslash{\mathcal{L}_b}})\wedge e(z_{\mathcal{L}_{\bar{\Theta}\oplus S}\backslash{\mathcal{L}_b}}))
\end{align} 
We consider the following three cases:
\begin{enumerate}[label=(\roman*)]
\item If $\beta<\epsilon_1$ and $\beta\le\frac{1}{2}$, Bob uniformly selects $n\beta$ coordinates from the erased coordinates and name the set of corresponding locations as $\mathcal{B}$.  In this case $Y_{\mathcal{L}_{\bar{\Theta}\oplus S}\backslash{\mathcal{L}_b}}=Y_{\mathcal{L}_\mathcal{B}}$
\begin{align}
\mathsf{Pr}&[\#(e(Z_{\mathcal{L}_{\bar{\Theta}\oplus S}\backslash{\mathcal{L}_b}}))\ge n(\beta\epsilon_3- \delta)]\ge\mathsf{Pr}[\#(e(Z_{\mathcal{L}_{\bar{\Theta}\oplus S}\backslash{\mathcal{L}_b}}))\ge n(\beta(\epsilon_2- \frac{\delta}{\beta}))]\nonumber\\
&\ge \mathsf{Pr}[\#(e(Z_{\mathcal{L}_{\bar{\Theta}\oplus S}\backslash{\mathcal{L}_b}}))= n\beta\wedge\#(e(Z_{\mathcal{L}_{\bar{\Theta}\oplus S}\backslash{\mathcal{L}_b}}))\ge n(\beta(\epsilon_2- \frac{\delta}{\beta}))]\nonumber\\
&\ge (1-K(n)\exp(-nD(\beta||\epsilon_1)))(1-K(n)\exp(-n\beta D(\epsilon_2- \frac{\delta}{\beta}||\epsilon_2)))\nonumber\\
&\ge 1-\xi_2\nonumber\\
\xi_2&\triangleq 2K(n)\exp(-n\min(D(\beta||\epsilon_1),\beta D(\epsilon_2- \frac{\delta}{\beta}||\epsilon_2)))
\end{align}

\item If $\epsilon_1< \beta<\frac{1}{2}$, Bob collects all erased coordinates and uniformly selects $n(\beta-\epsilon_1)$ coordinates from the non-erased coordinates, but not in $\mathcal{G}$ or $\mathcal{L}_\alpha$. Bob names the set of corresponding locations as $\mathcal{B}$. In this case $\mathcal{L}_{\bar{\Theta}\oplus S}\backslash{\mathcal{L}_b}=\mathcal{L}_\mathcal{B}\triangleq\mathcal{L}_{\mathcal{B}e}\cup\mathcal{L}_{\mathcal{B}\bar e}$ where $\mathcal{L}_{\mathcal{B}e}$ represents the coordinates in $\mathcal{L}_\mathcal{B}$, which are erased at Bob and similarly, $\mathcal{L}_{\mathcal{B}\bar e}$ represents the coordinates in $\mathcal{L}_\mathcal{B}$, which are non-erased at Bob
\begin{align}
\mathsf{Pr}&[\#(e(Z_{\mathcal{L}_\mathcal{B}}))\ge n(\beta\epsilon_3- \delta)]\nonumber\\
&\ge\mathsf{Pr}[\#(e(Z_{\mathcal{L}_\mathcal{B}}))\ge n(\epsilon_1\epsilon_2+(\beta-\epsilon_1)\epsilon_3-\delta)]\nonumber\\
&\ge\mathsf{Pr}[\#(e(Z_{\mathcal{L}_\mathcal{B}}))\ge n((\epsilon_1-\delta')(\epsilon_2-\delta’)+(\beta-\epsilon_1-\delta')(\epsilon_3- \delta’))]\nonumber\\
&\ge\mathsf{Pr}\left[\begin{array}{rcl}n(\epsilon_1-\delta')<|\mathcal{L}_{\mathcal{B}e}|&<&n(\epsilon_1+\delta')\wedge\\
\wedge\#(e(Z_{\mathcal{L}_{\mathcal{B}e}}))&\ge& |\mathcal{L}_{\mathcal{B}e}|(\epsilon_2-\delta')\wedge\\\wedge\#(e(Z_{\mathcal{L}_{\mathcal{B}\bar{e}}}))&\ge&|\mathcal{L}_{\mathcal{B}\bar e}|(\epsilon_3- \delta'))\end{array}\right]\nonumber\\
&\ge (1-K(n)\exp(-n\min(D(\epsilon_1-\delta'||\epsilon_1),D(\epsilon_1+\delta'||\epsilon_1))))\times\nonumber\\
&\quad\times(1-K(n)\exp(-n(\epsilon_1-\delta')D(\epsilon_2-\delta'||\epsilon_2))\times\nonumber\\
&\quad\times(1-K(n)\exp(-n(\beta-\epsilon_1-\delta')D(\epsilon_3-\delta'||\epsilon_3))\nonumber\\
&\ge 1-\xi_3\\
\delta'&\triangleq \frac{\delta}{4\max(\epsilon_1,\epsilon_2,\epsilon_3,\beta-\epsilon_1)}\\
\xi_3&\triangleq 3K(n)\exp(-n\Delta_1)\\
\Delta_1&\triangleq\min(D(\epsilon_1-\delta'||\epsilon_1),D(\epsilon_1+\delta'||\epsilon_1),(\epsilon_1-\delta')D(\epsilon_2-\delta'||\epsilon_2),(\beta-\epsilon_1-\delta')D(\epsilon_3-\delta'||\epsilon_3))
\end{align}

\item If $\beta>\frac{1}{2}$, Bob collects all the coordinates not in $\mathcal{G}$ or $\mathcal{L}_\alpha$  and uniformly selects $n(2\beta-1)$ coordinates from $\mathcal{G}$. Bob names the set of corresponding locations as $\mathcal{B}$. 
In this case, $\mathcal{L}_{\bar{\Theta}\oplus S}\backslash{\mathcal{L}_b}=\mathcal{L}_{\mathcal{B}e}\cup(\mathcal{L}_{\mathcal{B}\bar e}\backslash{\mathcal{L}_b})$.

    \begin{align}
\mathsf{Pr}&[\#(e(Z_{\mathcal{L}_{\bar{\Theta}\oplus S}\backslash{\mathcal{L}_b}}))\ge n(\beta\epsilon_3- \delta)]\nonumber\\
     &\ge   \mathsf{Pr}[\#(e(Z_{\mathcal{L}_{\bar{\Theta}\oplus S}\backslash{\mathcal{L}_b}}))\ge n(\epsilon_1\epsilon_2+(1-\epsilon_1-\beta)\epsilon_3-\delta)]\nonumber\\
&\ge\mathsf{Pr}[\#(e(Z_{\mathcal{L}_{\bar{\Theta}\oplus S}\backslash{\mathcal{L}_b}}))\ge n((\epsilon_1-\delta')(\epsilon_2-\delta’)+(1-\epsilon_1-\beta-\delta')(\epsilon_3- \delta’))]\nonumber\\
&\ge\mathsf{Pr}\left[\begin{array}{rcl}n(\epsilon_1-\delta')<|\mathcal{L}_{\mathcal{B}e}|&<&n(\epsilon_1+\delta')\wedge\\
\wedge\#(e(Z_{\mathcal{L}_{\mathcal{B}e}}))&\ge& |\mathcal{L}_{\mathcal{B}e}|(\epsilon_2-\delta')\wedge\\\wedge\#(e(Z_{\mathcal{L}_{\mathcal{B}\bar{e}}\backslash{\mathcal{L}_b}}))&\ge&|\mathcal{L}_{\mathcal{L}_{\mathcal{B}\bar e}\backslash{\mathcal{L}_b}}|(\epsilon_3- \delta'))\end{array}\right]\nonumber\\
&\ge (1-K(n)\exp(-n\min(D(\epsilon_1-\delta'||\epsilon_1),D(\epsilon_1+\delta'||\epsilon_1))))\times\nonumber\\
&\quad\times(1-K(n)\exp(-n(\epsilon_1-\delta')D(\epsilon_2-\delta'||\epsilon_2))\times\nonumber\\
&\quad\times(1-K(n)\exp(-n(1-\epsilon_1-\beta-\delta')D(\epsilon_3-\delta'||\epsilon_3))\nonumber\\
&\ge 1-\xi_4\\
\delta'&\triangleq \frac{\delta}{4\max(\epsilon_1,\epsilon_2,\epsilon_3,1-\epsilon_1-\beta)}\\
\xi_4&\triangleq 3K(n)\exp(-n\Delta_2)\\
\Delta_2&\triangleq\min(D(\epsilon_1-\delta'||\epsilon_1),D(\epsilon_1+\delta'||\epsilon_1),(\epsilon_1-\delta')D(\epsilon_2-\delta'||\epsilon_2),(1-\epsilon_1-\beta-\delta')D(\epsilon_3-\delta'||\epsilon_3))
\end{align}
\end{enumerate}
By defining $\xi_5=\min(\xi_2,\xi_3,\xi_4)$, we have
\begin{align}
\mathsf{Pr}&[\#(e(Z_{\mathcal{L}_{\bar{\Theta}\oplus S}\backslash{\mathcal{L}_b}}))\ge n(R- \delta)]\ge1-\xi_5
\end{align}
Therefore, we have
\begin{align}
    &H( F_{\bar{\Theta}\oplus S}(X_{\mathcal{L}_{\bar{\Theta}\oplus S}}) | X_{\mathcal{L}_{b}},\mathcal{L}_{b} , Z _{\mathcal{L}_{\bar{\Theta}\oplus S}}, F_{\bar{\Theta}\oplus S})\nonumber\\ &\geq 
 \mathsf{Pr}[\#(e(Z_{\mathcal{L}_{\bar{\Theta}\oplus S}\backslash{\mathcal{L}_b}}))\ge n(R- \delta)]\nonumber\\
 &\quad H( F_{\bar{\Theta}\oplus S}(X_{\mathcal{L}_{\bar{\Theta}\oplus S}}) | X_{\mathcal{L}_{b}},\mathcal{L}_{b} , Z _{\mathcal{L}_{\bar{\Theta}\oplus S}}, F_{\bar{\Theta}\oplus S},\#(e(Z_{\mathcal{L}_{\bar{\Theta}\oplus S}\backslash{\mathcal{L}_b}}))\ge n(R- \delta))\nonumber\\ &\geq
  (1-\xi_5)  \left(n(R-\tilde \delta) - \frac{2^{n(R-\tilde \delta)-n(R-\delta)}}{\ln2}\right)\nonumber\\ 
    &=  (1-\xi_5)  \left(n(R-\tilde \delta) - \frac{2^{-n(\tilde \delta -\delta)}}{\ln2} \right)\label{duse}
\end{align}
\end{enumerate}
For the third term, we have
\begin{align}
    I&(K_\Theta; \Theta,K_{\bar{\Theta}},V_E)\nonumber\\
    &=I(K_\Theta;\Theta,K_{\bar{\Theta}},Z^n,F)\nonumber\\
    &=I(K_\Theta;\Theta,K_{\bar{\Theta}},Z^n, \mathcal{L}_0, \mathcal{L}_1, \mathcal{L}_\alpha, S\oplus K_\alpha, F_\alpha, F_0, F_1,  K_{\Theta} \oplus F_{\Theta\oplus S}(X_{\mathcal{L}_{\Theta\oplus S}}), \nonumber\\&\quad \quad K_{\bar{\Theta}} \oplus  F_{\bar{\Theta}\oplus S}(X_{\mathcal{L}_{\bar{\Theta}\oplus S}}))\nonumber\\
    &\overset{(a)}{=}I(K_\Theta;\Theta,K_{\bar{\Theta}},Z^n,S, \mathcal{L}_0, \mathcal{L}_1, \mathcal{L}_\alpha, S\oplus K_\alpha, F_\alpha, F_0, F_1, K_{\Theta} \oplus F_{\Theta\oplus S}(X_{\mathcal{L}_{\Theta\oplus S}}), \nonumber\\&\quad \quad  K_{\bar{\Theta}} \oplus  F_{\bar{\Theta}\oplus S}(X_{\mathcal{L}_{\bar{\Theta}\oplus S}}))\nonumber\\
    &=I(K_\Theta;\Theta,K_{\bar{\Theta}},Z^n,S, \mathcal{L}_{\Theta\oplus S}, \mathcal{L}_{\bar{\Theta} \oplus S}, \mathcal{L}_\alpha,  K_\alpha, F_\alpha, F_{\Theta\oplus S}, F_{\bar{\Theta} \oplus S},  K_{\Theta} \oplus F_{\Theta\oplus S}(X_{\mathcal{L}_{\Theta\oplus S}}), \nonumber\\&\quad \quad F_{\bar{\Theta}\oplus S}(X_{\mathcal{L}_{\bar{\Theta}\oplus S}}))\nonumber\\
    &\overset{(b)}{=}I(K_\Theta;\Theta,Z^n,S, \mathcal{L}_{\Theta\oplus S}, \mathcal{L}_{\bar{\Theta} \oplus S}, \mathcal{L}_\alpha,  K_\alpha, F_\alpha, F_{\Theta\oplus S}, F_{\bar{\Theta} \oplus S},  K_{\Theta} \oplus F_{\Theta\oplus S}(X_{\mathcal{L}_{\Theta\oplus S}}),\nonumber\\&\quad \quad  F_{\bar{\Theta}\oplus S}(X_{\mathcal{L}_{\bar{\Theta}\oplus S}}))\nonumber\\
    &=I(K_\Theta;K_{\Theta} \oplus F_{\Theta\oplus S}(X_{\mathcal{L}_{\Theta\oplus S}}),  F_{\bar{\Theta}\oplus S}(X_{L_{\bar{\Theta}\oplus S}}) | \Theta,Z^n,S,\mathcal{L}_{\Theta\oplus S}, \mathcal{L}_{\bar{\Theta} \oplus S}, \mathcal{L}_\alpha,  K_\alpha, F_\alpha, F_{\Theta\oplus S}, F_{\bar{\Theta} \oplus S})\nonumber\\
    &=H(K_{\Theta} \oplus F_{\Theta\oplus S}(X_{\mathcal{L}_{\Theta\oplus S}}),  F_{\bar{\Theta}\oplus S}(X_{\mathcal{L}_{\bar{\Theta}\oplus S}}) | \Theta,Z^n,S, \mathcal{L}_{\Theta\oplus S}, \mathcal{L}_{\bar{\Theta} \oplus S}, \mathcal{L}_\alpha,  K_\alpha, F_\alpha, F_{\Theta\oplus S}, F_{\bar{\Theta} \oplus S})\nonumber\\&\quad -H(K_{\Theta} \oplus F_{\Theta\oplus S}(X_{\mathcal{L}_{\Theta\oplus S}}),  F_{\bar{\Theta}\oplus S}(X_{\mathcal{L}_{\bar{\Theta}\oplus S}}) |K_\Theta, \Theta,Z^n,S, \mathcal{L}_{\Theta\oplus S}, \mathcal{L}_{\bar{\Theta} \oplus S}, \mathcal{L}_\alpha,  K_\alpha, F_\alpha, F_{\Theta\oplus S}, F_{\bar{\Theta} \oplus S})\nonumber\\
    &\leq |F_{\Theta\oplus S}(X_{\mathcal{L}_{\Theta\oplus S}})| + |F_{\bar{\Theta}\oplus S}(X_{\mathcal{L}_{\bar{\Theta}\oplus S}})| \nonumber\\& \quad - H( F_{\Theta\oplus S}(X_{\mathcal{L}_{\Theta\oplus S}}),  F_{\bar{\Theta}\oplus S}(X_{\mathcal{L}_{\bar{\Theta}\oplus S}}) | K_\Theta,\Theta,Z^n,S, \mathcal{L}_{\Theta\oplus S}, \mathcal{L}_{\bar{\Theta} \oplus S}, \mathcal{L}_\alpha,  K_\alpha,  F_\alpha, F_{\Theta\oplus S}, F_{\bar{\Theta} \oplus S})\nonumber\\
    &\overset{(c)}{=} |F_{\Theta\oplus S}(X_{\mathcal{L}_{\Theta\oplus S}})| + |F_{\bar{\Theta}\oplus S}(X_{\mathcal{L}_{\bar{\Theta}\oplus S}})| \nonumber\\& \quad - H( F_{\Theta\oplus S}(X_{\mathcal{L}_{\Theta\oplus S}}),  F_{\bar{\Theta}\oplus S}(X_{\mathcal{L}_{\bar{\Theta}\oplus S}}) | Z_{\mathcal{L}_{\Theta\oplus S}},Z_{\mathcal{L}_{\bar{\Theta}\oplus S}}, \mathcal{L}_{\Theta\oplus S}, \mathcal{L}_{\bar{\Theta} \oplus S},  F_{\Theta\oplus S}, F_{\bar{\Theta} \oplus S})\nonumber\\
    &=|F_{\Theta\oplus S}(X_{\mathcal{L}_{\Theta\oplus S}})| + |F_{\bar{\Theta}\oplus S}(X_{\mathcal{L}_{\bar{\Theta}\oplus S}})| \nonumber\\& \quad - H( F_{\Theta\oplus S}(X_{\mathcal{L}_{\Theta\oplus S}}) | Z_{\mathcal{L}_{\Theta\oplus S}},Z_{\mathcal{L}_{\bar{\Theta}\oplus S}}, \mathcal{L}_{\Theta\oplus S}, \mathcal{L}_{\bar{\Theta} \oplus S},  F_{\Theta\oplus S}, F_{\bar{\Theta} \oplus S})\nonumber\\
    &\quad-H( F_{\bar{\Theta}\oplus S}(X_{\mathcal{L}_{\Theta\oplus S}}) | F_{\Theta\oplus S}(X_{\mathcal{L}_{\Theta\oplus S}}), Z_{\mathcal{L}_{\Theta\oplus S}},Z_{\mathcal{L}_{\bar{\Theta}\oplus S}}, \mathcal{L}_{\Theta\oplus S}, \mathcal{L}_{\bar{\Theta} \oplus S},  F_{\Theta\oplus S}, F_{\bar{\Theta} \oplus S})\nonumber\\
    &\overset{(d)}{=}|F_{\Theta\oplus S}(X_{\mathcal{L}_{\Theta\oplus S}})| + |F_{\bar{\Theta}\oplus S}(X_{\mathcal{L}_{\bar{\Theta}\oplus S}})| \nonumber\\& \quad - H( F_{\Theta\oplus S}(X_{\mathcal{L}_{\Theta\oplus S}}) |Z_{\mathcal{L}_{\Theta\oplus S}}, F_{\Theta\oplus S})\nonumber\\
    &\quad-H( F_{\bar{\Theta}\oplus S}(X_{\mathcal{L}_{\bar{\Theta}\oplus S}}) | F_{\Theta\oplus S}(X_{\mathcal{L}_{\Theta\oplus S}}), Z_{\mathcal{L}_{\Theta\oplus S}},Z_{\mathcal{L}_{\bar{\Theta}\oplus S}}, L_{\Theta\oplus S}, \mathcal{L}_{\bar{\Theta} \oplus S},  F_{\Theta\oplus S}, F_{\bar{\Theta} \oplus S})\nonumber\\ 
    &\leq |F_{\Theta\oplus S}(X_{\mathcal{L}_{\Theta\oplus S}})| + |F_{\bar{\Theta}\oplus S}(X_{\mathcal{L}_{\bar{\Theta}\oplus S}})| \nonumber\\& \quad - H( F_{\Theta\oplus S}(X_{\mathcal{L}_{\Theta\oplus S}}) | Z_{\mathcal{L}_{\Theta\oplus S}}, F_{\Theta\oplus S})\nonumber\\
    &\quad-H( F_{\bar{\Theta}\oplus S}(X_{L_{\bar{\Theta}\oplus S}}) | X_{L_{\Theta\oplus S}}, Z_{\mathcal{L}_{\Theta\oplus S}},Z_{\mathcal{L}_{\bar{\Theta}\oplus S}}, \mathcal{L}_{\Theta\oplus S}, \mathcal{L}_{\bar{\Theta} \oplus S},  F_{\Theta\oplus S}, F_{\bar{\Theta} \oplus S})\nonumber
\\
    &\overset{(e)}{=}|F_{\Theta\oplus S}(X_{\mathcal{L}_{\Theta\oplus S}})| + |F_{\bar{\Theta}\oplus S}(X_{\mathcal{L}_{\bar{\Theta}\oplus S}})| \nonumber\\& \quad - H( F_{\Theta\oplus S}(X_{\mathcal{L}_{\Theta\oplus S}}) | Z_{\mathcal{L}_{\Theta\oplus S}}, F_{\Theta\oplus S})-H( F_{\bar{\Theta}\oplus S}(X_{L_{\bar{\Theta}\oplus S}}) | X_{\mathcal{L}_{b}}, \mathcal{L}_{b}, Z_{\mathcal{L}_{\bar{\Theta}\oplus S}}, F_{\bar{\Theta} \oplus S})\nonumber\\ 
    &=2n(R-\tilde\delta)- H( F_{\Theta\oplus S}(X_{\mathcal{L}_{\Theta\oplus S}}) | Z_{\mathcal{L}_{\Theta\oplus S}}, F_{\Theta\oplus S},)-H( F_{\bar{\Theta}\oplus S}(X_{\mathcal{L}_{\bar{\Theta}\oplus S}}) | X_{\mathcal{L}_{b}}, \mathcal{L}_{b}, Z_{\mathcal{L}_{\bar{\Theta}\oplus S}}, F_{\bar{\Theta} \oplus S})\nonumber\\
    & \overset{(f)}{\leq} 2n(R-\tilde\delta)-(1-\xi_6)\left(  n(R-\tilde \delta) -\frac{2^{n(R-\tilde \delta)-n(R-\delta)}}{\ln2}\right)-(1-\xi_5)\left(  n(R-\tilde \delta) -\frac{2^{n(R-\tilde \delta)-n(R-\delta)}}{\ln2} \right)\nonumber\\
    &=(\xi_6+\xi_5) n(R-\tilde \delta) + (2-\xi_6-\xi_5) \frac{2^{-n(\tilde \delta -\delta)}}{\ln2}
\end{align}
where 
\begin{enumerate}[label=(\alph*)]
\item holds since $S$ is a function of $(\Theta,Z^n,\mathcal{L}_1,\mathcal{L}_0)$, 
\item holds since $K_{\bar{\Theta}}$ is independent of all other random variables, 
\item holds because of the following Markov chain
\begin{align} 
(F_{\Theta\oplus S}(X_{\mathcal{L}_{\Theta\oplus S}}),  F_{\bar{\Theta}\oplus S}(X_{\mathcal{L}_{\bar{\Theta}\oplus S}}) &\rightarrow (Z_{\mathcal{L}_{\Theta\oplus S}},Z_{\mathcal{L}_{\bar{\Theta}\oplus S}}, \mathcal{L}_{\Theta\oplus S}, \mathcal{L}_{\bar{\Theta} \oplus S},  F_{\Theta\oplus S}, F_{\bar{\Theta} \oplus S})\nonumber\\
 &\rightarrow (K_\Theta,\Theta,Z^n,S, \mathcal{L}_\alpha, F_\alpha(X_{\mathcal{L}_\alpha}), F_\alpha)
\end{align} 
\item holds because of the following Markov chain
\begin{align} (F_{C\oplus S}(X_{\mathcal{L}_{\Theta\oplus S}}) \rightarrow (Z_{\mathcal{L}_{\Theta\oplus S}}, F_{\Theta\oplus S}) \rightarrow (Z_{\mathcal{L}_{\bar{\Theta}\oplus S}}, \mathcal{L}_{\Theta\oplus S}, \mathcal{L}_{\bar{\Theta} \oplus S}, F_{\bar{\Theta} \oplus S})
\end{align} 
\item holds because of the following Markov chain 
\begin{align} (F_{\bar{\Theta}\oplus S}(X_{\mathcal{L}_{\bar{\Theta}\oplus S}})) \rightarrow (X_{\mathcal{L}_{b}}, Z_{\mathcal{L}_{\bar{\Theta}\oplus S}}, \mathcal{L}_{b}, F_{\bar{\Theta} \oplus S}) \rightarrow (Z_{\mathcal{L}_{\Theta\oplus S}}, \mathcal{L}_{\Theta\oplus S}, \mathcal{L}_{\bar{\Theta} \oplus S},  F_{\Theta\oplus S},X_{\mathcal{L}_{C\oplus S}})
\end{align}  
\item is due to the following derivation. We note that $\mathcal{L}_{\Theta\oplus S}=\mathcal{G}$
\begin{align}
    R(X_{\mathcal{L}_{\mathcal{G}}}|Z_{\mathcal{L}_{\mathcal{G}}}=z_{\mathcal{L}_{\mathcal{G}}}) &=\#(e(z_{\mathcal{L}_{\mathcal{G}}}))
\end{align}
 where $\#(e(z_{\mathcal{L}_{\mathcal{G}}}))$ represents the number of coordinates, at which $z_{\mathcal{L}_{\mathcal{G}}}$ are erased. We note that
 \begin{align}
 \mathsf{Pr}&[\#(e(Z_{\mathcal{L}_{\mathcal{G}}})) \geq n(R-\delta)]
 \ge\mathsf{Pr}[\#(e(Z_{\mathcal{L}_{\mathcal{G}}})) \geq n(\beta\epsilon_3-\delta)]\nonumber\\
 &\ge \mathsf{Pr}[|\mathcal{L}_{\mathcal{G}}|= n\beta\wedge\#(e(Z_{\mathcal{L}_{\mathcal{G}}}))\ge n(\beta(\epsilon_3- \frac{\delta}{\beta}))]\nonumber\\
&\ge (1-K(n)\exp(-nD(\beta||(1-\epsilon_1))))(1-K(n)\exp(-n\beta D(\epsilon_3- \frac{\delta}{\beta}||\epsilon_3)))\nonumber\\
&\ge 1-\xi_6\\
\xi_6&\triangleq 2K(n)\exp(-n\min(D(\beta||(1-\epsilon_1)),\beta D(\epsilon_3- \frac{\delta}{\beta}||\epsilon_3)))
 \end{align}
 
By using Lemma~\ref{lem:privacy_amplification}, we have
\begin{align}
    &H(F_{\Theta\oplus S}(X_{\mathcal{L}_{\Theta\oplus S}})| Z_{\mathcal{L}_{\Theta\oplus S}},F_{\Theta\oplus S})\nonumber\\ &\geq 
 \mathsf{Pr}[\#(e(Z_{\mathcal{L}_{\mathcal{G}}})) \geq n(R-\delta)] H(F_{\Theta\oplus S}(X_{\mathcal{L}_{\Theta\oplus S}})| Z_{\mathcal{L}_{\Theta\oplus S}},F_{\Theta\oplus S},\#(e(Z_{\mathcal{L}_{\mathcal{G}}})) \geq n(R-\delta))   \nonumber\\ &\geq
  (1-\xi_6)  \left(n(R-\tilde \delta) - \frac{2^{n(R-\tilde \delta)-n(R-\delta)}}{\ln2}\right)\nonumber\\ 
    &=  (1-\xi_6)  \left(n(R-\tilde \delta) - \frac{2^{-n(\tilde \delta -\delta)}}{\ln2} \right)
\end{align}
From (\ref{duse}), we have 
\begin{align}
H(F_{\bar{\Theta}\oplus S}(X_{\mathcal{L}_{\bar{\Theta}\oplus S}})| X_{\mathcal{L}_b}, L_b,Z_{\mathcal{L}_{\bar{\Theta} \oplus S}},F_{\bar{\Theta} \oplus S}) \geq  (1-\xi_5)\left(n(R-\tilde \delta) -\frac{2^{-n(\tilde \delta -\delta)}}{\ln2}\right)\end{align}
\end{enumerate}

\section{Proof of Corollary \ref{cor_pro}}\label{pro_cor}
In BESBC, we have 
\begin{align}
C_0&=1\\
C_G&=\epsilon_3\\
C_B&=\epsilon_2-1
\end{align}
Since $C_B\le 0$, we have that the optimal $\tau_1,\tau_2$ are
\begin{align}
\tau_1&=0\\
\tau_2&=\gamma_1-\gamma_2
\end{align}
Then, the lower bound in Theorem \ref{Theo:general lower bound} is simplified to
\begin{align}
R=\max_{\gamma_1,\gamma_2}\min\left\{\begin{array}{l}
(\gamma_1-\gamma_2)\\
\gamma_2\epsilon_3+(\gamma_1-\gamma_2)\epsilon_3\\
\gamma_2\epsilon_3+(\gamma_1-\gamma_2)\epsilon_2\\
\frac{1}{2}[\min(\gamma_1+\gamma_2,(1-\epsilon_1))\epsilon_3 +(\gamma_1-\gamma_2)\epsilon_2]
\end{array}\right.\label{low_bin}
\end{align}
where parameters $\gamma_1,\gamma_2$ satisfy
\begin{align}
0\le\gamma_1&\le1-\epsilon_1\label{feas1}\\
0\le\gamma_1-\gamma_2&\le\epsilon_1\label{feas2}
\end{align}
We note that the above lower bound is a linear programming problem. Therefore, the optimal $\gamma_1,\gamma_2$ should take the value of the corner points of the feasible set specified by (\ref{feas1}) and (\ref{feas2}). Among all the corner points, we observe that the optimum should be 
\begin{align}
\gamma_1=1-\epsilon_1,\gamma_1-\gamma_2=\epsilon_1&\quad\text{ if }\epsilon\le\frac{1}{2}\label{corn1}\\
\gamma_1=1-\epsilon_1,\gamma_1-\gamma_2=1-\epsilon_1&\quad\text{ if }\epsilon\ge\frac{1}{2}\label{corn2}
\end{align}
and the lower bound in (\ref{low_bin}) is simplified to
\begin{align}
R=\min\left\{\begin{array}{l}
\min(\epsilon_1,1-\epsilon_1)\\
(1-\epsilon_1)\epsilon_3\\
\max(1-2\epsilon_1,0)\epsilon_3+\min(\epsilon_1,1-\epsilon_1)\epsilon_2\\
\frac{1}{2}[(1-\epsilon_1)\epsilon_3 +\min(\epsilon_1,1-\epsilon_1)\epsilon_2]
\end{array}\right.\label{low_bin_fi}
\end{align}
If we assume $\epsilon_2\ge\epsilon_3$, as in Theorem \ref{Theo:lower bound}, we note that the second term in (\ref{low_bin}) is no larger than the third term in the lower bound in (\ref{low_bin}), then we have 
\begin{align}
R= \min\left(\min(\epsilon_1,1-\epsilon_1), (1-\epsilon_1)\epsilon_3, \frac{1}{2}[(1-\epsilon_1)\epsilon_3 +\min(\epsilon_1,1-\epsilon_1)\epsilon_2]\right)
\end{align}
We note that if $\epsilon_1\le\frac{1}{2}$, then
\begin{align}
R\le \min\left(\epsilon_1, (1-\epsilon_1)\epsilon_3,\frac{1}{2}[(1-\epsilon_1)\epsilon_3+\epsilon_1\epsilon_2]\right)
\end{align}
If $\epsilon_1\ge\frac{1}{2}$
\begin{align}
\min&\left(\min(\epsilon_1,1-\epsilon_1), (1-\epsilon_1)\epsilon_3, \frac{1}{2}[(1-\epsilon_1)\epsilon_3 +\min(\epsilon_1,1-\epsilon_1)\epsilon_2]\right)\nonumber\\
&=\min\left(1-\epsilon_1, (1-\epsilon_1)\epsilon_3, \frac{1}{2}[(1-\epsilon_1)(\epsilon_3 +\epsilon_2)]\right)\nonumber\\
&=(1-\epsilon_1)\epsilon_3\nonumber\\
&=\min\left(\epsilon_1, (1-\epsilon_1)\epsilon_3,(1-\epsilon_1)\epsilon_2\right)\nonumber\\
&=\min\left(\epsilon_1, (1-\epsilon_1)\epsilon_3,\epsilon_1\epsilon_2\right)\nonumber\\
&=\min\left(\epsilon_1, (1-\epsilon_1)\epsilon_3,\frac{1}{2}[(1-\epsilon_1)\epsilon_3+\epsilon_1\epsilon_2]\right)
\end{align}
which is the same as Theorem \ref{Theo:lower bound}.

If we assume $\epsilon_2<\epsilon_3$, we have that the second term is larger than the third term in the lower bound in (\ref{low_bin}). Therefore, 
\begin{align}
R&= \min\left(\min(\epsilon_1,1-\epsilon_1), \max(1-2\epsilon_1,0)\epsilon_3+\min(\epsilon_1,1-\epsilon_1)\epsilon_2, \frac{1}{2}[(1-\epsilon_1)\epsilon_3 +\min(1-\epsilon_1,\epsilon_1)\epsilon_2]\right)\nonumber\\
&=\left\{\begin{array}{ll}
\min\left(\epsilon_1, (1-2\epsilon_1)\epsilon_3+\epsilon_1\epsilon_2, \frac{1}{2}[(1-\epsilon_1)\epsilon_3 +\epsilon_1\epsilon_2]\right)&\text{ if }\epsilon_1\le\frac{1}{2}\\
(1-\epsilon_1)\epsilon_2&\text{ if }\epsilon_1\ge\frac{1}{2}
\end{array}\right.
\end{align}
which concludes the proof of Corollary \ref{cor_pro}.

\bibliographystyle{IEEEtran}
\bibliography{ref}

\end{document}